\documentclass[conference]{IEEEtran}
\usepackage{cite}
\usepackage{amsmath,amssymb,amsfonts}
\usepackage{algorithmic}
\usepackage[linesnumbered, ruled]{algorithm2e}
\usepackage{graphicx}
\usepackage{subfigure}
\usepackage{float}
\usepackage{booktabs}
\usepackage{cite}
\usepackage{textcomp}
\usepackage{xcolor}
\usepackage{booktabs}
\usepackage{fancyhdr}
\usepackage{import}
\usepackage{url}
\usepackage[T1]{fontenc}
\usepackage[top = 0.6 in,left =0.6in,right=0.6in, bottom=0.6in, includefoot]{geometry}
\def\BibTeX{{\rm B\kern-.05em{\sc i\kern-.025em b}\kern-.08em    T\kern-.1667em\lower.7ex\hbox{E}\kern-.125emX}}

\begin{document}

\title{A Distributed Learned Hash Table}
\author{
\IEEEauthorblockN{Shengze Wang\textsuperscript{1}, Yi Liu\textsuperscript{1}, Xiaoxue Zhang\textsuperscript{2}, Liting Hu\textsuperscript{1}, Chen Qian\textsuperscript{1}}
\textit{\textsuperscript{1}University of California Santa Cruz, \textsuperscript{2}University of Nevada Reno}
\\
\{shengze, yliu634, liting, cqian12\}@ucsc.edu, xiaoxuez@unr.edu
}
\maketitle

\thispagestyle{plain}

\begin{abstract}
 Distributed Hash Tables (DHTs) are pivotal in numerous high-impact key-value applications built on distributed networked systems, offering a decentralized architecture that avoids single points of failure and improves data availability. Despite their widespread utility, DHTs face substantial challenges in handling range queries, which are crucial for applications such as LLM serving, distributed storage, databases, content delivery networks, and blockchains. To address this limitation, we present LEAD, a novel system incorporating learned models within DHT structures to significantly optimize range query performance. LEAD utilizes a recursive machine learning model to map and retrieve data across a distributed system while preserving the inherent order of data. LEAD includes the designs to minimize range query latency and message cost while maintaining high scalability and resilience to network churn. Our comprehensive evaluations, conducted in both testbed implementation and simulations, demonstrate that LEAD achieves tremendous advantages in system efficiency compared to existing range query methods in large-scale distributed systems, reducing query latency and message cost by 80\% to 90\%+. Furthermore, LEAD exhibits remarkable scalability and robustness against system churn, providing a robust, scalable solution for efficient data retrieval in distributed key-value systems. 
\end{abstract}
\section{Introduction}
Key-value data management across distributed computing systems plays a crucial role in supporting large-scale Internet applications, including the emerging area of large language model (LLM) serving~\cite{wang2024towards,vllm,sglang24,kdn,CacheGen2024,arxiv.2504.20101}.
Distributed Hash Tables (DHTs) have been widely used for decentralized data management~\cite{stoica2001chord,zarrin2018resource,passarella2012survey}.
A DHT is a distributed data structure adept at performing storage and retrieval operations of key-value pairs across a decentralized network of nodes. DHTs mitigate the limitations of centralized architectures by eliminating single points of failure and distributing data loads across numerous nodes, thereby enhancing data availability and network efficiency~\cite{coluzzi2023survey}. State-of-the-art systems like InterPlanetary File System (IPFS)~\cite{trautwein2022design}, Cassandra~\cite{Apache, lakshman2010cassandra}, Tor~\cite{torprojectProjectPrivacy}, Namecoin~\cite{Namecoin}, and Bittorrent~\cite{BitTorrent}, have exemplified the integration of DHTs in ensuring scalable and fault-tolerant data management
in distributed networked systems.

\textbf{The problem.} Despite their widespread adoption and inherent advantages, DHT-based systems encounter significant challenges, particularly when handling complex queries such as range queries, which are important functions in applications such as KV-cache sharing in LLM serving~\cite{vllm,sglang24,srivatsa2024preble,kdn}, distributed file systems and databases~\cite{trautwein2022design, Apache}, edge-cloud systems\cite{Fog-Chiang,IoTDI19}, and blockchain systems~\cite{MARIJAN2022100492,nie2024collaborative}. Current  DHT systems are primarily optimized for single-key lookups.
DHTs use a uniformly random hash function to distribute keys into random locations, hence similar keys will be mapped to completely different storage locations. This feature of DHT will introduce two major limitations for range queries. First, all keys in the queried range need to be searched to ensure the completeness of the query. Second, these keys will be mapped to different locations based on the hash function. Accessing these locations will cause a high cost of network traffic. 
In the literature, efforts to improve range query performance in distributed systems have led to limited solutions. Armada ~\cite{4527242} uses a partition tree model within the FissionE topology~\cite{1498449}. DBST~\cite{9642540} integrates binary search trees for range queries. 
MARQUES~\cite{7043516} employs space-filling curves in a multi-level overlay structure, bringing increased overhead and scalability issues. RQIOT~\cite{djellabi2020effective} explores the idea of using order-preserving hashing to improve range query efficiency, yet how to design such a hash method, especially in a dynamic distributed system, is unclear. These solutions cannot completely resolve the two limitations of range queries in DHT. 

\textbf{Our solution.} To address the critical issue -- enabling efficient range queries for distributed networked systems -- we introduce \textbf{LEAD} (LEArned DHT), a novel system that first integrates machine learning models with DHT frameworks to enhance the performance of range queries evidently. Drawing on the learned indexes proposed in recent years~\cite{kraska2018case}, which suggests that indexes could be conceptualized as "models" that predict the position of a key within a dataset, we argue that \textbf{a learned model can replace the hash function to distribute keys in networked systems}. By learning the cumulative distribution function (CDF) of keys, we can \textbf{maintain the inherent order of these keys} while mapping them to a decentralized group of nodes, making similar keys be placed in close locations. Hence the two limitations of random hash functions can be completely resolved. To minimize inference overhead and reduce the prediction error, we adapt the Recursive Model Index (RMI) structure~\cite{10.1145/3318464.3384706} to train the learned model. 

However, the idea of learning models to maintain the key relationships while disturbing keys consistently in DHT-based systems poses several challenges. First, we need to devise a strategy for managing key mapping and peer addressing, as well as utilizing the relationships between keys to conduct range queries efficiently. Second, the distributed environment is highly dynamic and characterized by frequent network churns; this requires the protocol to quickly adapt to network changes. Third, as the network expands and new data are introduced, the previously established Cumulative Distribution Function (CDF) on which the model was trained may no longer accurately represent the new data distribution. Consequently, the learned model might not distribute data as uniformly as traditional hash functions, posing additional challenges for load balancing.

In response to these challenges, our protocol, LEAD, elaborates on the methodologies for applying learned models within DHT-based systems, focusing on the following aspects:

\begin{itemize}
\item[(1)] We first introduce the concept of the Learned Hash Function under the realm of distributed key-value systems. We detail the strategy to map and retrieve keys with learned models for DHT-based systems. This approach renovates traditional hash functions that map keys to random positions, allowing LEAD to maintain the inherent order of keys and enhance range query performance.
\item[(2)] LEAD is designed to adapt dynamically to frequent changes in the system such as database size increases, node joins, and departures. It employs mechanisms that rapidly update the overlay routing tables and maintain the learned models, ensuring the system remains robust and efficient even in highly volatile environments. We propose a distributed model update method termed the Federated Recursive Model (FRM). 
\item[(3)] LEAD incorporates a load-balancing model called \textit{Shadow Balancer} using virtual nodes to allocate keys in an even manner that prevents overloading specific nodes, thus enhancing overall system performance and scalability.
\item[(4)] We conduct comprehensive evaluations of LEAD's performance in both implementation on real networked systems and simulations.  The evaluation spans various network conditions, scales, and topologies, along with diverse datasets and data volumes. Our assessment demonstrates LEAD significantly outperforms existing baseline methods in range query efficiency, reducing latency by more than tenfold compared to traditional methods in current DHT-based systems. Additionally, LEAD exhibits remarkable scalability and resilience to network churn, maintaining logarithmic efficiency in single-key query performance. 
\item[(5)] We conduct two timely case studies demonstrating LEAD's effectiveness in real-world applications that require efficient range queries: key-value cache management for LLM serving and the InterPlanetary File System (IPFS).
\end{itemize}

Beyond the immediate motivation of accelerating range queries in classical DHT deployments, the same order‑preserving learned hash that powers LEAD unlocks a diverse set of emerging workloads: it can collocate semantically close embeddings in vector databases that serve retrieval‑augmented LLMs, shard the rapidly growing key–value caches and adapter weights of distributed transformer inference without a central router, deliver geo‑temporal IoT telemetry and edge‑AI models to nearby gateways for low‑latency analytics, adapt CDN object placement to shifting popularity skew in real time, and provide an range index across heterogeneous blockchains. These broader scenarios underscore LEAD’s potential as a general storage substrate for next‑generation, data‑intensive distributed networked systems and motivate the design choices detailed in the rest of the paper. 

\section{Background and Motivation}
\label{sec:background}
\begin{figure}[t]
    \centering
    \setlength{\subfigcapskip}{-8pt}
    \subfigure[Memory cost]{
        \begin{minipage}[t]{0.2\textwidth}
            \includegraphics[width=3.5cm]{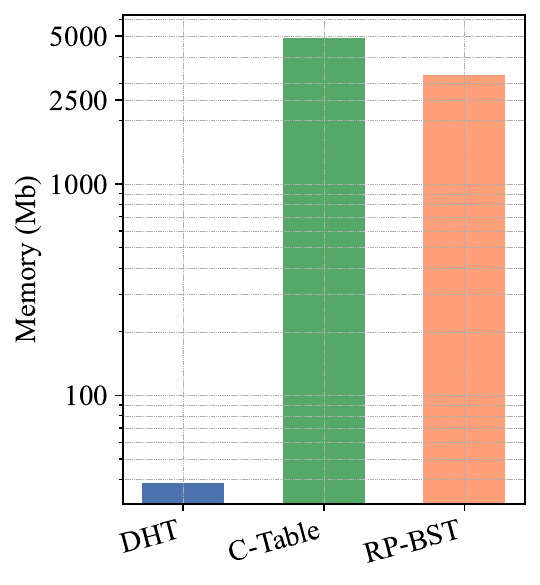} 
        \end{minipage}
        \label{m1}
     }
    \subfigure[\# of messages]{
        \begin{minipage}[t]{0.2\textwidth}
        \includegraphics[width=3.5cm]{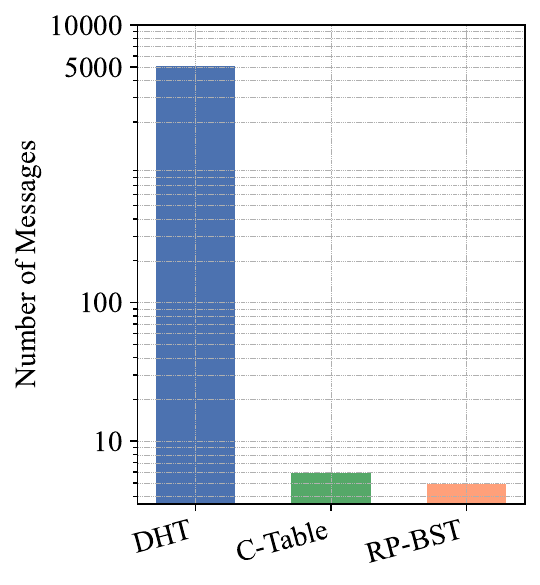}
        \end{minipage}
    \label{m2}
    }
    \vspace{-2ex}
    \caption{Micro benchmark of Range Query Performance}
    \label{micro}
    \vspace{-4ex}
\end{figure}
\label{section:micro}
We conducted a focused micro‑benchmark to expose the inherent trade‑offs of three representative baselines: DHT-based key-value system with Chord~\cite{stoica2001chord} (DHT), the centralized range-location mapping table (C-Table), and the Range‑Partition Binary Search Tree (RP‑BST), whose details are presented in Appendix\ref{sec:benchmark}. Chord utilizes a ring-like hashing space to manage key-value pairs and is highly efficient for single-key lookups due to its logarithmic routing efficiency. However, it struggles with range queries, which often require traversing multiple nodes sequentially, thereby increasing latency and message cost. We also implemented DBST~\cite{9642540} as an RP‑BST overlay. Each node maintained a BST interval and two routing pointers (left/right). We evaluated the number of messages required to complete range queries and their memory overhead — critical metric affecting response time and the efficiency of data retrieval in distributed environments. The experimented system includes 100 nodes with 200 million key-value pairs from the `osmc64’ dataset (described in Section \ref{dataset}) and executed range queries for a range covering 2,000 keys after a given key.  As depicted in Fig. \ref{micro}, the centralized table substantially reduces the number of messages required to resolve range queries compared to DHT; however, it imposes a higher memory burden on the system. RP-BST–style overlays also improve messaging efficiency; however, they incur considerable memory consumption and control-plane complexity. Crucially, they offer limited resilience to network churn and impose high costs for index maintenance. Furthermore, both the centralized table and RP-BST overlay require a dedicated coordinator to maintain and synchronize metadata, introducing an additional bottleneck in distributed deployments. This underscores the necessity for a solution like LEAD, which aims to merge the advantages of both solutions. 
 
\begin{figure*}[htb]
	\begin{minipage}[t]{0.35\linewidth}
		\centering
		\includegraphics[width=2.5in]{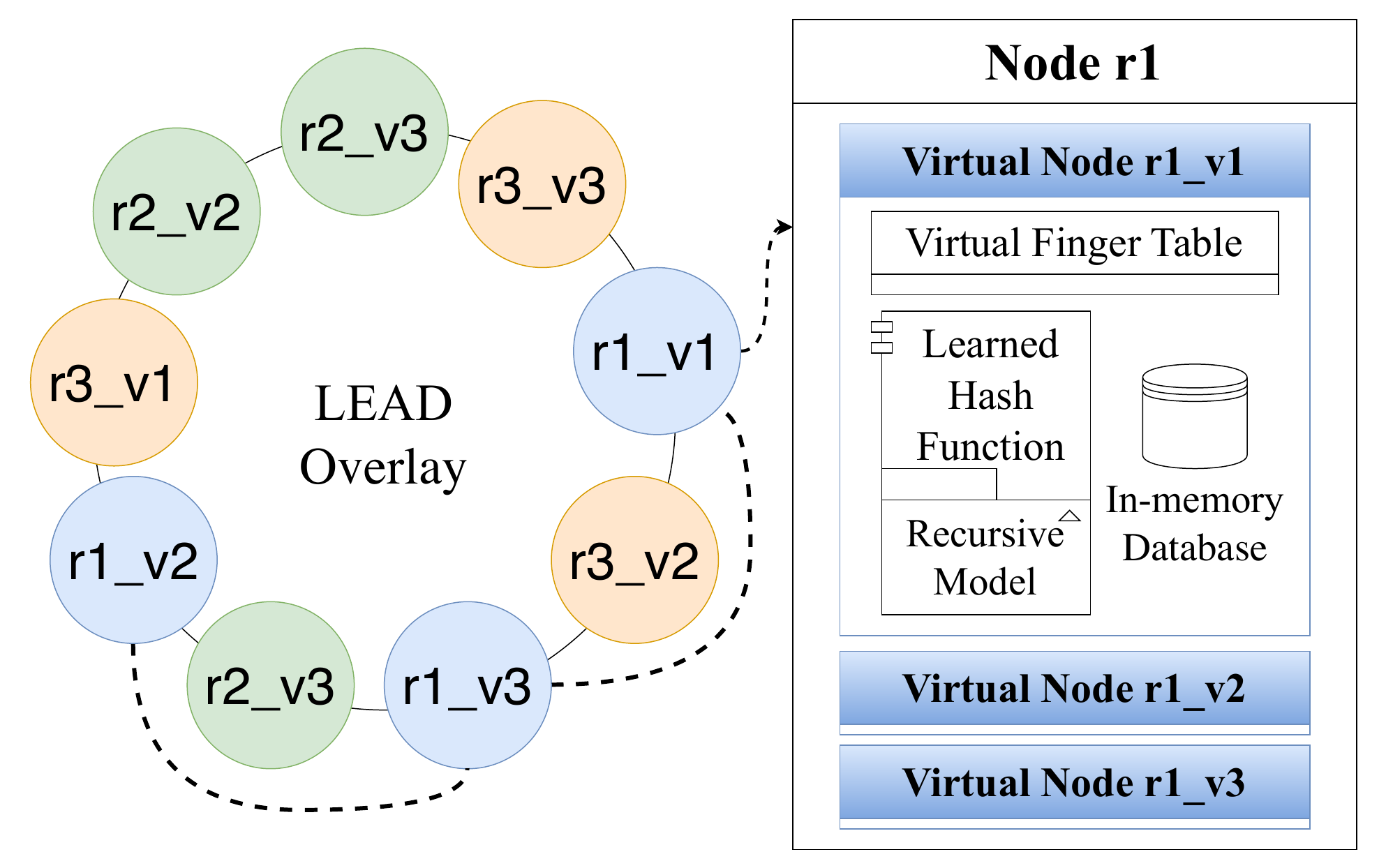}
        \vspace{-1ex}
		\caption{LEAD System Design}
		\label{s1}
	\end{minipage}
	\begin{minipage}[t]{0.35\linewidth}
		\centering
		\includegraphics[width=2.5in]{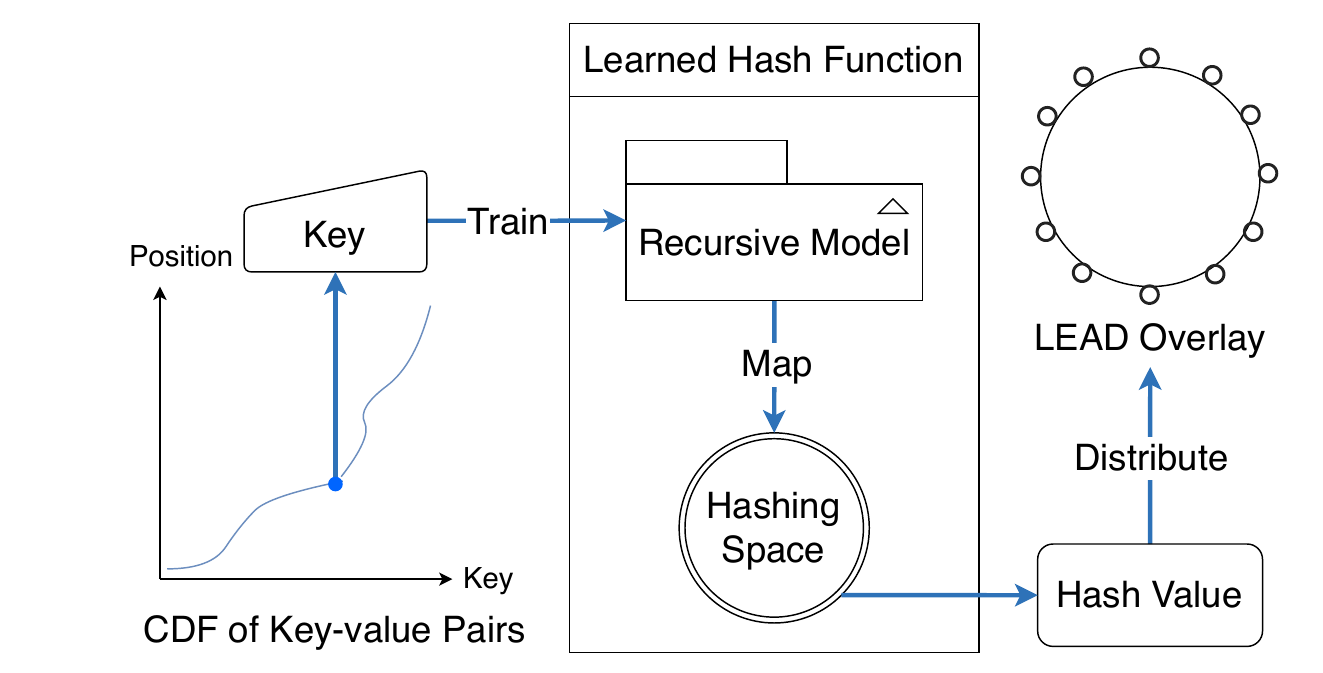}
        \vspace{-1ex}
		\caption{Key mapping with a learned hash function}
		\label{s2}
	\end{minipage}
    \begin{minipage}[t]{0.29\linewidth}
		\centering
		\includegraphics[width=2in]{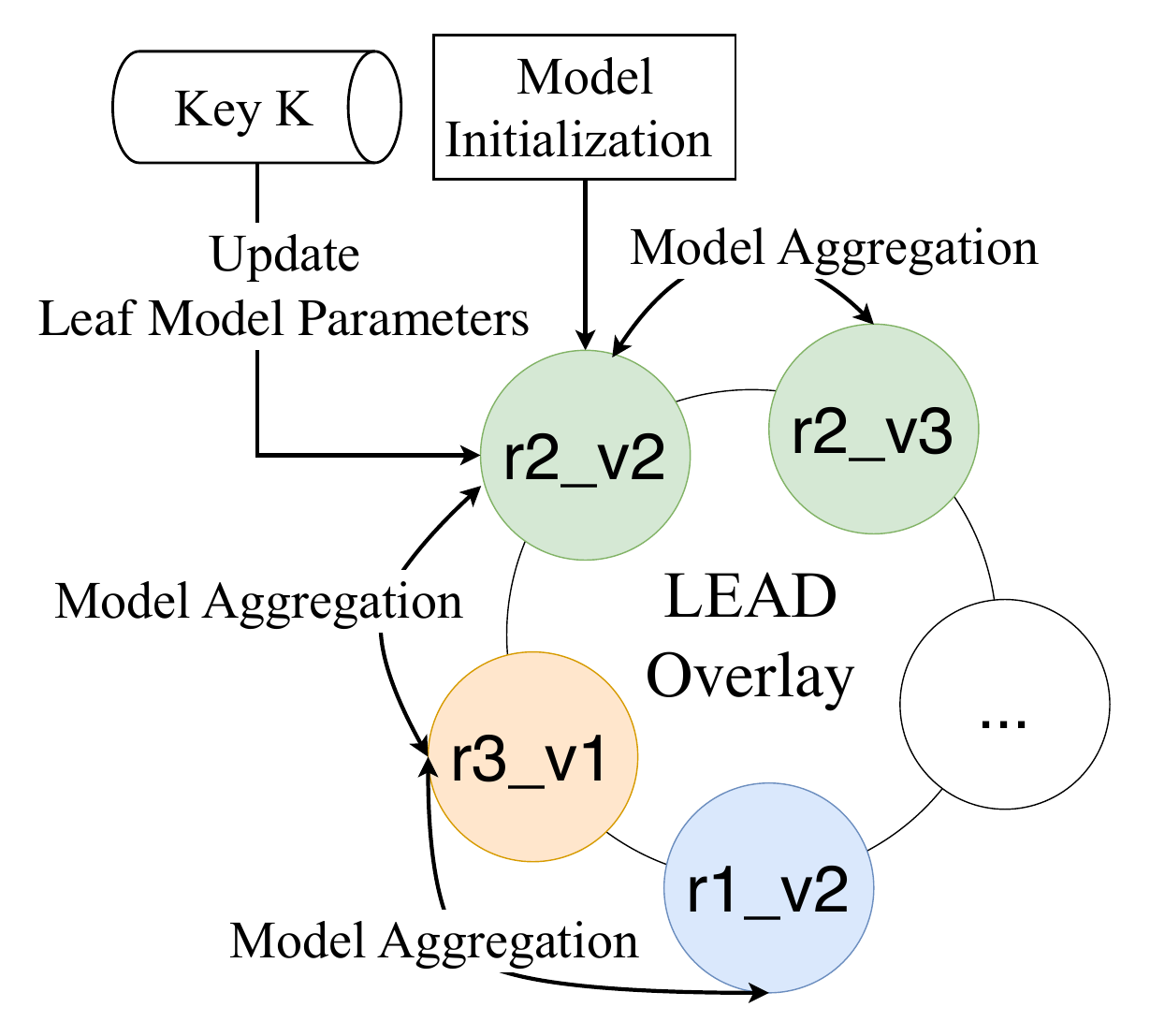}
        \vspace{-1ex}
		\caption{Decentralized Model Update}
		\label{s4}
	\end{minipage}
    \vspace{-3ex}
\end{figure*}
\section{LEAD Design}
\vspace{-1ex}
\label{sec:design}
\subsection{System Overview}
The section describes the LEAD system. It details the methodologies employed in LEAD for key mapping using its \textit{Learned Hash Function}, addressing peers during node joins and departures, data retrieval mechanisms tailored for queries, and balancing the loads. Additionally, the protocol outlines stabilization and recovery strategies to handle system dynamics.

Fig.~\ref{s1} presents the system design of LEAD. At a high level, physical nodes within the system are virtualized into multiple virtual nodes, each functioning as independent peers within a structured overlay network. Central to each peer is the learned model utilized for efficient and in-order key mapping. This is complemented by a consistent hashing function employed specifically for peer addressing. Each peer also maintains a virtual finger table, the component for storing updated routing information and facilitating effective data queries. Additionally, peers are equipped with an in-memory database dedicated to the storage and rapid retrieval of key-value pairs.

\vspace{-1ex}
\subsection{Key Mapping with a Learned Hash Function}
\label{train1}
\vspace{-1ex}
LEAD uses a learned hash function for key mapping, as showed in Fig.~\ref{s2}. 
Unlike traditional hash functions, which aim to map keys to random values within a specified range, the learned hash function strategically maps keys to order-preserving values in a hashing space. Utilizing the cumulative distribution function (CDF) of keys managed on the network, it maintains the inherent order of these keys while mapping them to a hashing space. This preservation of key relationships enhances systems with the capability for in-order data retrieval. 

We employ the Recursive Model Indexes (RMI) structure~\cite{10.1145/3318464.3384706} to implement the learned hash function in LEAD. In Section \ref{sec:models}, we will show that RMI provides the lowest latency compared to other learned models. The RMI structure is a hierarchy of models, where at each stage the model determines the appropriate child model to engage for a specified key.  At the leaf level, models predict the relative position of a key within a dataset. A scale factor, $S$, is then applied to translate this relative position into a hashing space comprising $H$ hash values. For instance, considering a two-stage RMI trained on $N$ key-value pairs, the learned hash function, denoted as $LearnedHASH$, can be articulated as follows:
\vspace{-5ex}
\\ \hspace*{\fill} \\
\begin{minipage}{.47\textwidth}
\begin{equation}
LearnedHASH(key) = \lfloor \frac{N}{H} \times f_2^{\lfloor \frac{B\footnote{B referred to as the branching factor that determimines the number of "buckets" that data is divided into by the stage-one model} \times f_1(x)\footnote{$f_i$ referred to as the \textit{i}th stage model}}{N}\rfloor}(K)\rfloor
\end{equation}
\vspace{-25pt}
\end{minipage}
\\ \hspace*{\fill} \\
$LearnedHASH$ is trained by optimizing the parameters of the given model by minimizing the squared error of its predictions. Specifically, a model $k$ at stage $\rho$, denoted by $f_\rho^{(k)}$, is trained with the following loss function~\cite{kraska2018case}:
\begin{minipage}{.47\textwidth}
\begin{equation}
L_\rho = \sum_{(x,y)\footnote{$x$ is the key, $y \in [0,N)$ is its relative position within a dataset}}^{}(f_\rho^{\lfloor \frac{M_\rho\footnote{ Number of models at stage $\rho$} \times f_{\rho - 1}(x)}{N} \rfloor}(x) - y)^2
\end{equation}
\vspace{-14pt}
\end{minipage}
\\ \hspace*{\fill} \\
\vspace{-3ex}

\textbf{We introduce three systems‑level optimizations that are critical for a fully‑decentralized overlay for the vanilla RMI:} (i) Auto‑Model Selection. At bootstrap time a peer runs a lightweight mountain‑climbing probe adapted from the learned‑index tuner —that trains candidate leaf predictors on a 1\% sketch of its local key sample, ranks them by 99‑th‑percentile prediction error and instantiates the for an optimal trade-off between model size and prediction error. (ii) The vanilla RMI assumes a fixed target domain. In practice, node joins and virtual‑node churn change the effective density of the overlay, so a leaf that once mapped to may need only half that span an amount of updates later. Each peer therefore attaches a 2‑field anchor 〈offset,scale〉 to its leaf model: the on‑line gradient update adjusts offset to keep the median key centered and dials scale up/down with a 2‑bit PID controller so the 95 \% key‑quantile always ends near the right edge of the peer’s virtual‑ID window.  (iii) LEAD integrates a Federated Recursive Model (FRM) within its Learned Hash Function, enabling collaborative learning among peers for dynamic model updates. This decentralized design ensures load balancing and seamless request handling during model updates, as detailed in Section \ref{section:update}. $LearnedHASH$ maps each key to a hash value within the same hashing space used for peer addressing. While hash collisions for different keys are permissible, the hash value's primary role is to distribute the key across the network, not to serve as a unique identifier. Each key-value pair, identified by the key $K$, is assigned to the first peer whose $VID$ (as detailed in Section \ref{peer-add}) either equals or follows the hash value produced by $LearnedHASH(K)$. 

\textbf{Model initialization and re-training.} We assume the system starts with a small number (<10) of nodes with a limited amount of data. Hence, the very first model training can be conducted on an arbitrary node without causing a scalability problem. Then more nodes and data join the system, hence \textbf{one of the key contribution of LEAD is to adjust the network for newly joined nodes and re-train the learned model for new data}. The re-training mechanism will be detailed in Sec.~\ref{section:update}.
\vspace{-0.5ex}
\subsection{Load balancing with virtual nodes}
\vspace{-0.5ex}
Achieving balanced load distribution in distributed key-value systems remains challenging. These systems contain heterogeneous nodes, with varied storage capacity and network bandwidth. Additionally, nodes may experience resource shortages due to higher-priority tasks or hotspots (popular data items that attract many requests). These factors undermine the randomization and uniformity that consistent hashing aims for, leading to uneven load distribution, bottlenecks, and inefficiencies within the system. To address these challenges, LEAD employs a load balancing model called \textit{Shadow Balancer}, which utilizes virtual nodes to optimize key distribution across the network and alleviate hotspot effects. As illustrated in Fig. \ref{s1}, each physical node is virtualized into multiple virtual nodes, with each operating as an independent peer within the network. To facilitate efficient peer addressing and data retrieval processes, this design also leverages consistent hashing to ensure that these virtual peers are distributed as evenly as possible across the hashing space. The operational policy of the \textit{Shadow Balancer} is formalized as follows:
\begin{itemize}
\item[(1)] Each node virtualizes itself into 
$k$ virtual peers, where 
$k$ is adjustable according to the node's capabilities.
\item[(2)] In response to resource bottlenecks, a node plans the departure of virtual peers that manage fewer requests. 
\end{itemize}
Even in resource-constrained environments, the Shadow Balancer adds minimal overhead. See Appendix~\ref{sec:balancer_analysis} for its detailed analysis.
\vspace{-1ex}
\subsection{Peer Addressing\label{peer-add}}
\vspace{-1ex}
Along with the learned hash function, LEAD employs a consistent hashing mechanism known as $PeerHASH$ to assign an $m$-bit identifier, denoted as $VID$, to each peer in the network. Specifically, our implementation of LEAD utilizes a universal hash function as $PeerHASH$. Each physical node, referred to as $N$, hosts one or more virtual nodes, collectively called $V$. These virtual nodes are assigned unique port numbers, enabling direct inter-peer communication without intermediaries. The $VID$ for each peer can be derived by hashing a concatenation of the corresponding node’s IP address and its port number using $PeerHASH$. Every $V$ maintains its own set of network routing information in a structure known as \textit{virtual finger table}.
In a hashing space holding $h$ hashing values, the table holds $\lfloor \log h \rfloor$ entries, with each entry comprising a $VID$ and the corresponding node's IP address. Similar to Chord,  each $i^{th}$ entry in the virtual finger table of a virtual node $V$ identifies the first node, $S$, that succeeds $V$ by at least $10^{i-1}$ positions in the hashing space for peer addressing. We defines the $Successor(x)$ as the first peer whose $VID$ is equal to or follows a hash value $x$ in the peer addressing space. Consequently, the \textit{i}th entry of the virtual finger table of $V$, denoted as $vfinger(i)$, can be formalized as
\vspace{-5ex}
\\ \hspace*{\fill} \\
\begin{minipage}{.47\textwidth}
\begin{equation}
vfinger(i) = successor (VID + 10^{i-1})
\end{equation}
\end{minipage}
\vspace{-3ex}
\\ \hspace*{\fill} \\

\vspace{-1ex}
\subsubsection{Node Joins and Departs}
To maintain the status of $V$ in a dynamic network, each peer $V$ must preserve the status of its successor. The process for a node ($N$) to join the network is outlined in the following procedures:
\begin{itemize}
\item[(1)] \textbf{Initialization:} A new node initializes itself either as the first node in an empty network or by obtaining information about an existing peer ($V_0$) that is part of the network.
\item[(2)] \textbf{Node virtualization:} The node $N$ creates $n$ virtual nodes ($V$s) and assigns them $n$ unique ports. Their Virtual IDs ($VID$s) are then generated using the $PeerHASH$.
\item[(3)] \textbf{Successor Discovery:}  Each virtual node $V$ dispatches a Remote Procedure Call (RPC) to $V_0$ to lookup for $Successor(VID)$ and obtain its knowledge of the network, including the successor's predecessor, successor, and virtual finger table. The lookup mechanics for $Successor(VID)$ are further detailed in Section \ref{look-up}.
\item[(4)] \textbf{Status Acknowledge:} Upon identifying its successor peer $Successor(VID)$, the virtual node $V$ establishes itself within the network by setting $Successor(VID)$ as its immediate successor and adopting $Successor(VID)$'s current predecessor as its own. Subsequently, $V$ issues RPC to $Successor(VID)$ instructing it to update its predecessor record to $V$. Concurrently, $V$ sends another RPC to the predecessor of $Successor(VID)$, requesting an update of its successor record to $V$. Moreover, $V$ copies the 2nd to $i^{th}$ entries of the virtual finger table from $Successor(VID)$, which accelerates its initialization and stabilizes its initial operations within the network.
\item[(5)] \textbf{Key Transfer:} Once the virtual node $V$ has successfully joined the network, it initiates the key transfer process: $V$ requests $Successor(VID)$ to transfer the appropriate key-value pairs that fall within its responsibility range. 
\end{itemize}

For planned node departures, the node $N$ notifies the immediate successors and predecessors of its managed virtual nodes $V$. Subsequently, these virtual nodes $V$ transfer their key-value pairs to their predecessors.

\subsubsection{Virtual Finger Table Update}
Accurate and up-to-date routing information is crucial for the efficiency and reliability of LEAD. LEAD maintains the peer addressing information in virtual finger tables. Periodically, each peer updates its virtual finger table by sending RPCs across the network to obtain each entry's latest successor and their status. Additionally, events such as node joins, departures, and failures trigger the affected nodes to update their virtual finger tables.
\vspace{-1.5ex}
\subsection{Data Retrieval}
\label{retr}
\vspace{-1ex}
\subsubsection{Single Key Lookup\label{look-up}}
The distributed single key lookup process in LEAD aims to locate the immediate successor of a key by identifying the first peer on the network whose VID equals or follows the hash value of the given key in the hashing space. 
$P$ consults its virtual finger table to execute an optimal jump towards the key's hash identifier. This involves selecting the farthest preceding peer in the finger table that does not exceed the key's identifier, assuming this peer possesses closer or direct knowledge of the key. The query is then routed to this selected node, which follows the same procedure.
This iterative process continues until the query reaches the peer responsible for managing the key, denoted as $S$. Upon locating the key, $S$ dispatches an RPC directly back to $P$ with the requested data, effectively completing the retrieval process with enhanced efficiency and minimized latency.


\subsubsection{Range query\label{range-query}}
\begin{figure}[t]
\centering
\includegraphics[width=0.8\linewidth]{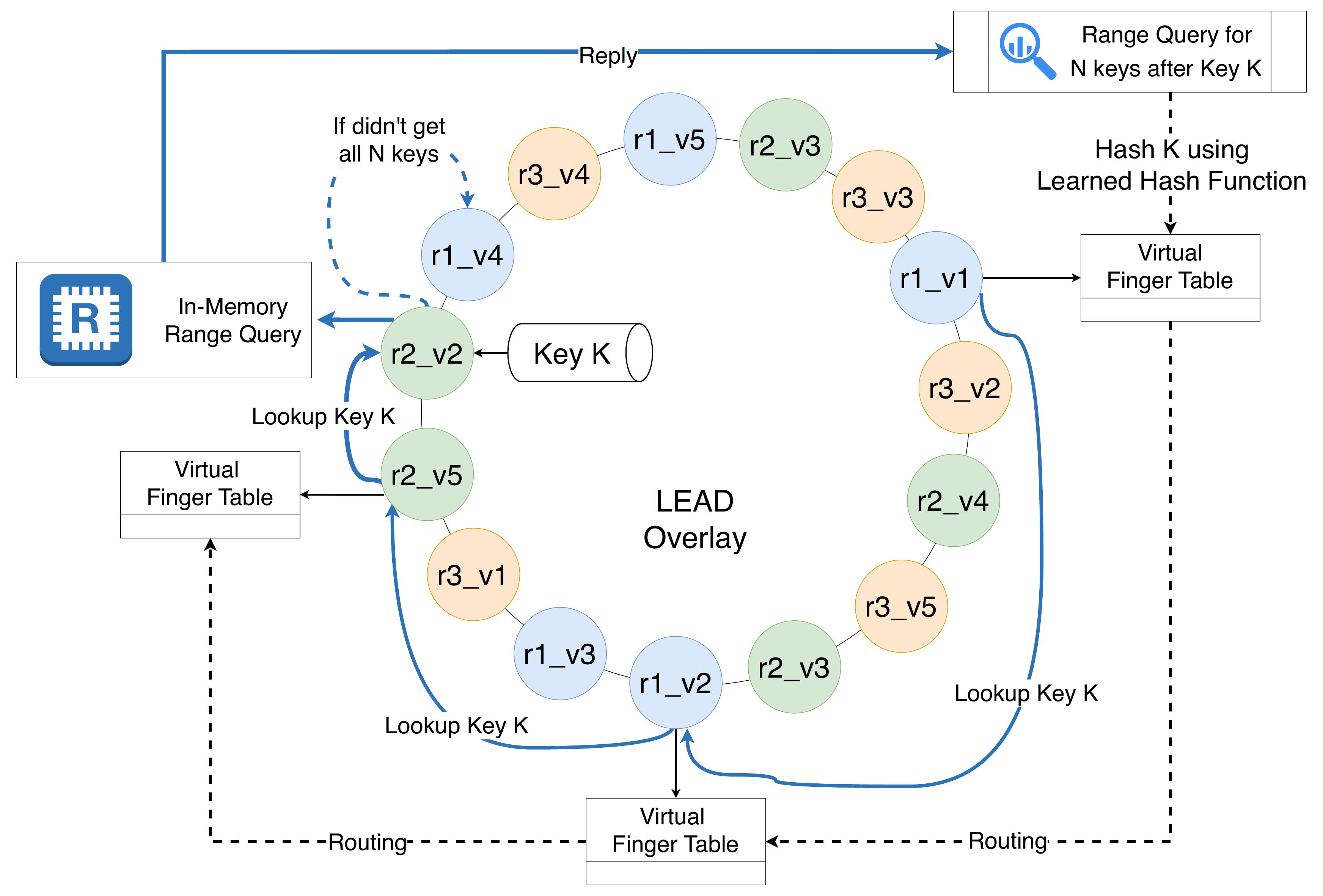} 
\vspace{-2.5ex}
\caption{Range Query in LEAD}
\vspace{-5ex}
\label{s3}
\end{figure}

LEAD leverages the Learned Hash Function to distribute keys across the network while preserving their relationships in order-preserving hash values. Range queries in LEAD are handled based on the order-preserving key mapping by the Learned Hash Function. To execute a range query for a sequence of $n$ successive key-value pairs starting from key $K$, the initiating peer $P$ first applies the Learned Hash Function to hash $K$ (as shown in Fig. \ref{s3}), yielding the hash value $L_K$. Using the single-key lookup mechanism described in Section \ref{look-up}, $P$ locates the peer $S$ responsible for $K$. Once the query reaches $S$, $S$ performs a local range query within its in-memory database to retrieve the sequence of key-value pairs. 
If $S$ holds only a portion of the required sequence, it forwards the remaining query to its successor. This forwarding process is repeated, moving through the chain of successors, until all $n$ keys are retrieved. The final peer to fulfill the range query then sends the complete set of results back to the initiating peer $P$.


\subsubsection{Model Update}
\label{section:update}
While the learned hash function in LEAD efficiently distributes new key-value pairs across the network, challenges arise when this model no longer aligns with the overall Cumulative Distribution Function (CDF) of the keys managed across the network. Such misalignment can lead to increased hash collisions and an uneven distribution of key-value pairs, potentially overloading specific network peers. As detailed in Section \ref{update1}, the Learned Hash does not necessitate updates until new key-value pairs constitute up to 40\% of the network’s storage for the tested datasets. \textbf{Sub-optimized learned hash functions do not impact the correctness of system operations}, but they may affect load balancing if there is a significant logarithmic discrepancy between the learned hash function and the current data distribution. Model updates can help optimize the workload balancing across the network. To effectively manage these discrepancies, LEAD is proposed with the Federated Recursive Model (FRM) within its Learned Hash Function, promoting decentralized and cooperative learning among peers for dynamic model updates. As showed in Fig.~\ref{s4}, FRM incorporates the hierarchical structure of Recursive Models, with each peer in the network incrementally refining its segment of the leaf models based on locally observed data changes. The L0 layer in the FRM structure performs approximate predictions to identify the leaf model for specific keys. The hierarchical structure maintains stability in the L0 parameters when the model captures the approximate CDF of existing data. As such, when new keys are integrated into the network, the focus of FRM is on refining the corresponding leaf models for the unlearned keys. Each peer operates with two versions of the Recursive Model: one active in the current Learned Hash Function and another reserved for updates. The system continuously monitors key distribution across the network in a decentralized manner through the tracking of the proportion of new key-value pairs integrated since the last model update at the peer level. When a new key-value pair is introduced to the network, the corresponding peer calculates the median index of keys it manages to determine the relative index for training, using its copy of the model designated for updates. The peer then selects the appropriate leaf model based on L0 layer predictions. Once the leaf model is identified, the peer refines this model. The relative index for training each new key is calculated by determining the median index of immediate keys currently managed by the peer. Given a network comprising $n$ peers, with $k$ keys distributed through the Learned Hash Function, we explore the scenario where $m$ additional keys are introduced. To ascertain the proportion of these new keys observed by any given peer causes the total new keys on the network to exceed a predefined threshold $t$, we can model this expectation as $\frac{m}{k+m}$, assuming a relatively balanced load across the network. Then, we can achieve the threshold at $t$ of the new key-value pairs observed by a peer, where the total new keys on the network exceed $t$ of the total keys managed on the network since the last update with high probability. 
During the early phase of the LEAD network, when only a few peers are present, a randomly selected training peer is designated to initialize FRM. This initial coordinator is selected based on criteria such as computational power and network load. Once chosen, all peers in the network transfer their key-value pairs to this node. The central node then performs batch training to establish the initial parameters for the learned hash function. The process begins with a lightweight Model‑Scout module that benchmarks multiple candidate leaf families (e.g., Linear, RadixSpline). A quick mountain-climbing search is then used to tune the model parameters, aiming for an optimal trade-off between model size and prediction error. 
Upon successful training, the model is adopted by other peers on the network through the stabilization process as discussed below. Peers are actively monitoring the proportion of new key-value pairs joined since the last model update. Once the proportion of new key-value pairs observed exceeds a threshold - specifically, 40\% as identified in our empirical study in Section \ref{update1}, the peer flags the readiness status for the model update as true in its heartbeat message. Upon a peer being ready for a model update and detecting that a majority of its neighbors on the successor and predecessor list (e.g., 90\%) are also flagged for updates, it takes the role of a transient coordinator. Then, it sends the flagged neighbors a Remote Procedure Call (RPC) to request confirmation of status and transfer of parameters. When such RPC is received by a peer, it pushes the updated leaf parameters to the transient coordinator, acknowledges readiness for the model update, and then resets its update-ready status, ensuring no redundant or conflicting update processes occur. During the parameter transfer, \textbf{only the segments that have changed are pushed to minimize data transfer size—for instance, only about 12 KB for approximately 1000 linear leaf model parameters and their segments stored in 32-bit format}. After receiving acknowledgments from its neighbors, the transient coordinator aggregates the updated leaf model parameters from these peers through the averaging operation. Once the new model is consolidated, a new version number will be assigned to facilitate network-wide recognition and adoption. Peers in LEAD periodically check for the latest model version via heartbeat messages with their neighbors. The sectional transient inconsistency caused by updates does not compromise the continuous service of the LEAD system, as peer-addressing relies on an independent hash function. Moreover, \textbf{during cooperative model updates, the system remains operational}; only a subset of peers performs asynchronous updates on the leaf models. This is targeted at specific key segments and occurs until significant data changes are detected. Thus, the integrity of the system is preserved.

\vspace{-1ex}
\subsection{Stabilization and Failures Recovery}
\label{stable}  
Handling system churn -- where nodes frequently join or depart -- is crucial for sustaining system integrity and performance. LEAD is designed to adapt rapidly to these changes through robust stabilization and failure recovery mechanisms. The correctness of LEAD is dependent on the current knowledge of its successors and predecessors within the network. Additionally, the efficiency of query handling is contingent upon the timeliness and accuracy of the virtual finger tables. To maintain this information, each peer periodically stabilizes themselves in the network through 
successor and predecessor verification, heartbeat communications, and virtual finger table maintenance.
Building on its stabilization mechanisms, LEAD incorporates resilient failure recovery strategies to address peer failures. The details are presented in the Appendix~\ref{sec:recovery}.
\section{Evaluation}
\label{sec:eva}

\begin{figure*}[ht]
    \centering
    \setlength{\subfigcapskip}{-5pt}
    \begin{minipage}[b]{0.99\textwidth}
    \centering
    \includegraphics[width=1\textwidth]{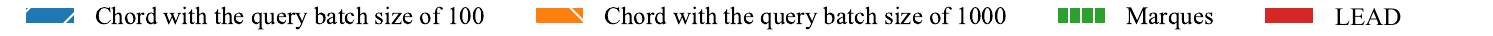} 
    \end{minipage}
    
    \subfigure[osmc64]{
        \begin{minipage}[b]{0.23\textwidth}
            \includegraphics[width=1\textwidth]{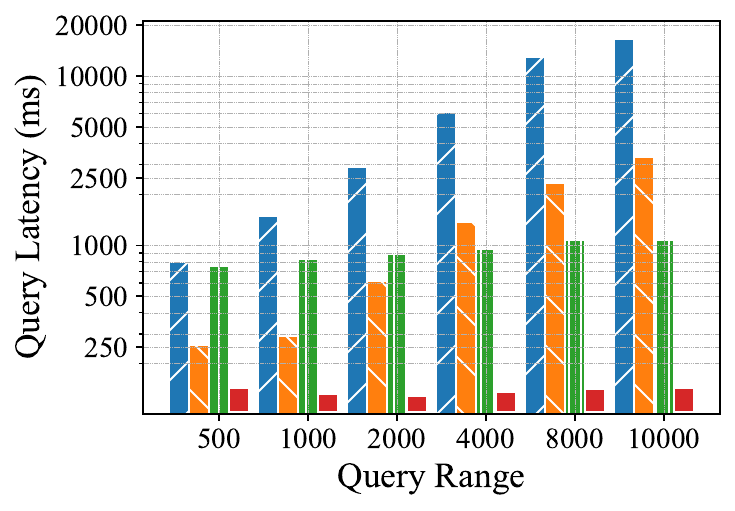} 
        \end{minipage}
        \label{real1}
    }
    \subfigure[face64]{
        \begin{minipage}[b]{0.23\textwidth}
        \includegraphics[width=1\textwidth]{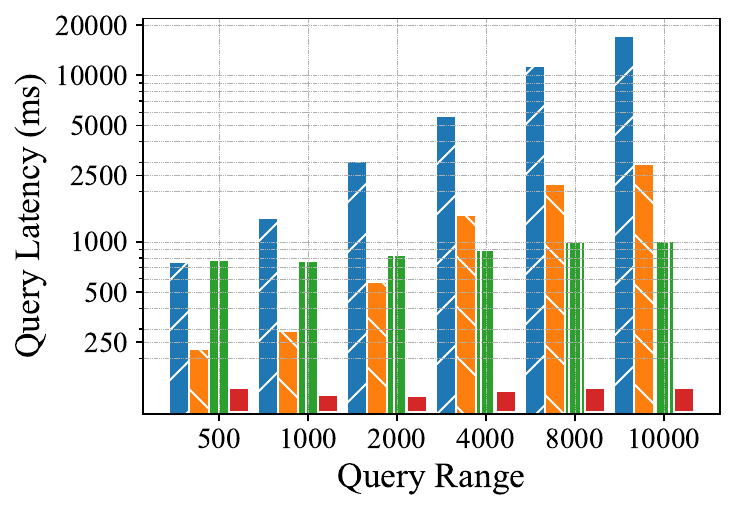}
        \end{minipage}
    \label{real2}
    }
    \subfigure[amzn64]{
        \begin{minipage}[b]{0.23\textwidth}
            \includegraphics[width=1\textwidth]{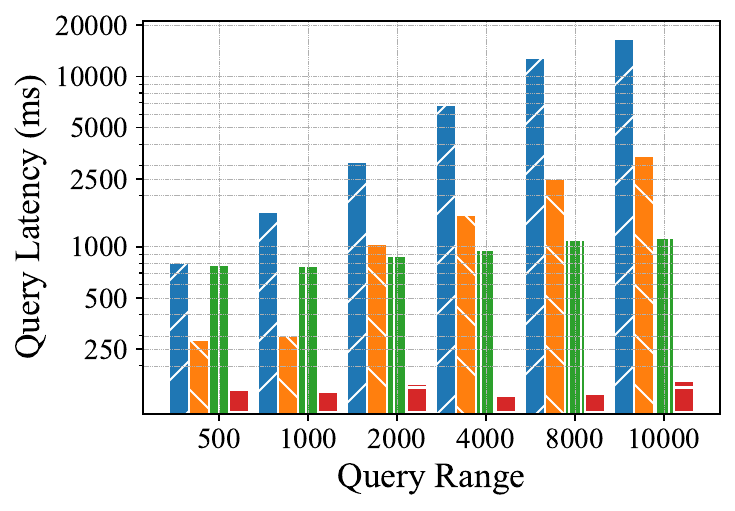} 
        \end{minipage}
        \label{real3}
    }
    \subfigure[wiki64]{
        \begin{minipage}[b]{0.23\textwidth}
        \includegraphics[width=1\textwidth]{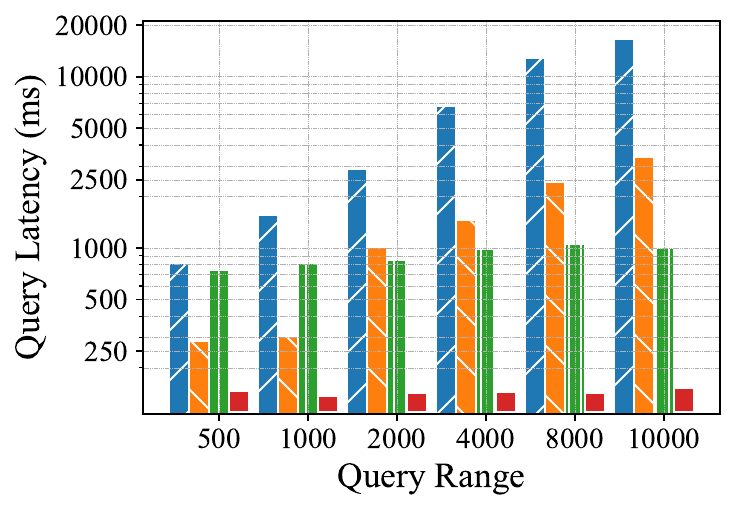}
        \end{minipage}
    \label{real4}
    }
    \vspace{-1ex}
    \caption{Latency of range queries on various datasets in the real-machine testbed}
     \vspace{-2ex}
    \label{real}
\end{figure*}

This section presents the evaluation of LEAD through both testbed implementation and large-scale simulations, along with real-world case studies.

\vspace{-1ex}
\subsection{Methodology}

\noindent\textbf{Hardware and environments.}
The testbed implementation comprises nine virtual machines in public clouds, including three types of machines: one with two Intel Xeon Silver 4314 2.40 GHz 16-Core CPUs and 128GB of DDR4 2666MHz memory; one with an Intel Xeon E5-2687W v4 3.00GHz 12-Core CPU and 32GB of DDR4 2400MHz memory; and the other with an Intel Core i7-7700 3.60GHz 4-Core CPU and 16GB of DDR4 2400MHz memory. They communicate through the Internet. Each virtual machine runs 10 virtual nodes in the overlay hence the overlay includes 90 peers in total. We utilize Redis for in-memory key-value storage on peers. 

The simulator we built, called p2psim+, is based on a publicly-available discrete event-driven simulator p2psim \cite{gil2003p2psim} running on an Ubuntu 22.04 LTS desktop with an AMD Ryzen 7 3700X 3.6 GHz 8-Core CPU, complemented by 32GB of DDR4 3200MHz RAM across two 16GB modules. P2psim is widely recognized and utilized within the community 
\cite{risson2007survey}. 
We added over 3,000 lines of C++ code to enhance the simulator. These extensions include the integration of LEAD, the support for user-defined network topologies, customized network behavior observers, and scalability enhancements for large experiments. We utilize the implementation of RMIs in Rust~\cite{10.1145/3318464.3384706}. 
We will publish p2psim+ upon the acceptance of this paper. 

\noindent\textbf{Datasets.\label{dataset}}
We leverage four real-world datasets from the SOSD benchmark~\cite{kipf2019sosd}, each consisting of 200 million 64-bit unsigned integers as keys. The datasets encapsulate a broad spectrum of data distributions and sources, described as follows:
\vspace{-1ex}
\begin{itemize}
\item[(1)] `osmc64': uniformly sampled OpenStreetMap Cell IDs
\item[(2)] `face64': randomly sampled Facebook user IDs
\item[(3)] `amzn64':  Amazon book sale popularity data
\item[(4)] `wiki64': Wikipedia article edit timestamps
\end{itemize}
\vspace{-1ex}
To accurately emulate real-world network topologies in our simulations, we incorporate the PlanetLab Dataset from the Network Latency Datasets \cite{7604140}. This dataset captures round-trip times (RTTs) between 490 nodes dispersed across the PlanetLab network. Specifically, we employ the "PlanetLabData\_1" as the latency model to construct the PlanetLab topology.

\noindent\textbf{Baselines.}
We use four baseline methods in our experiments: the batch query approach on Chord~\cite{stoica2001chord} DHT with batch sizes of either 100 or 1000, and the recent work Marques~\cite{7043516}. 
We let Chord batch single-key queries together and send them as one or multiple consolidated requests across the network.
Marques~\cite{7043516} is a recent enhancement on Chord~\cite{stoica2001chord} for range query efficiency. We exclude DBST~\cite{9642540} from direct comparison, as it relies on a centrally constructed binary search tree and incurs high overhead—analyzed in Section~\ref{section:micro}—that renders it unsuitable for decentralized environments. Similarly, RQIOT~\cite{djellabi2020effective} assumes centralized order-preserving hashing without providing a decentralized construction mechanism. Neither DBST nor RQIOT offer open-source implementations, further limiting their applicability in reproducible and fair comparison within our distributed system framework.

\vspace{-1ex}
\subsection{Testbed Performance\label{real-world}}
\begin{figure*}[htbp]
    \centering
    \setlength{\subfigcapskip}{-5pt}
    \begin{minipage}[b]{0.99\textwidth}
    \includegraphics[width=1\textwidth]{Figures/legend1.pdf} 
    \end{minipage}
    
    \subfigure[osmc64]{
        \begin{minipage}[b]{0.23\textwidth}
            \includegraphics[width=1\textwidth]{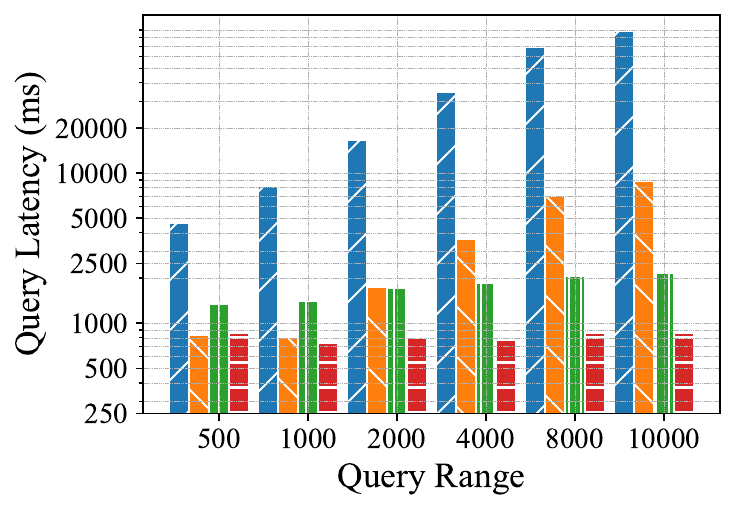} 
        \end{minipage}
        \label{sim_r1}
    }
    \subfigure[face64]{
        \begin{minipage}[b]{0.23\textwidth}
        \includegraphics[width=1\textwidth]{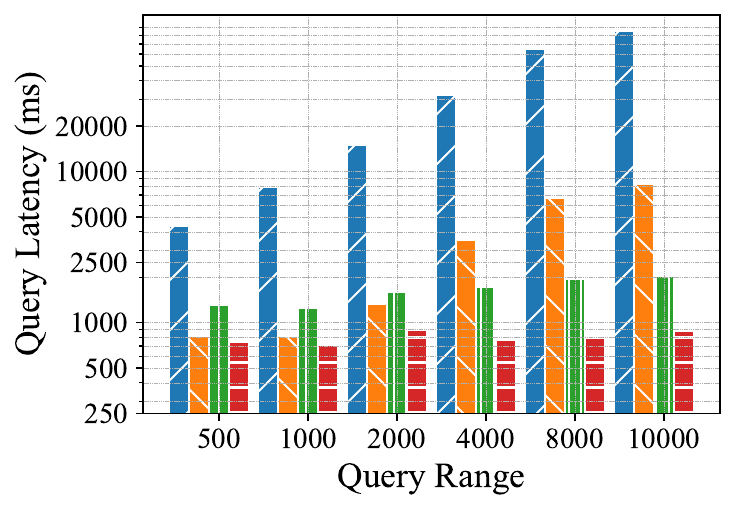}
        \end{minipage}
    \label{sim_r2}
    }
    \subfigure[amzn64]{
        \begin{minipage}[b]{0.23\textwidth}
            \includegraphics[width=1\textwidth]{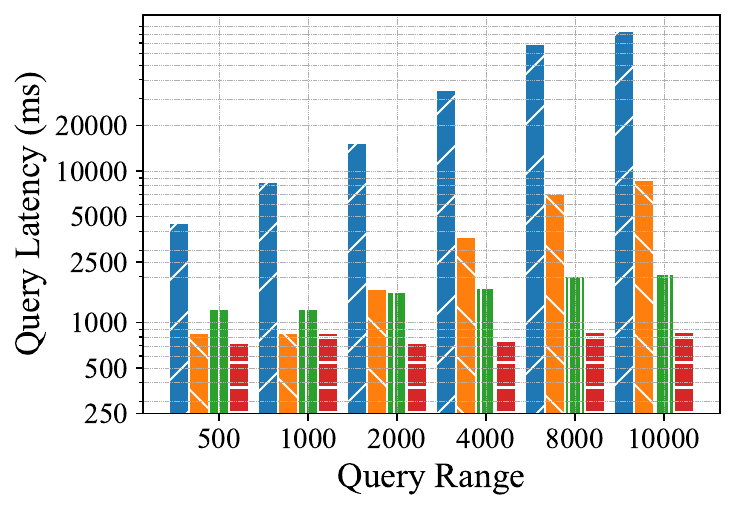} 
        \end{minipage}
        \label{sim_r3}
    }
    \subfigure[wiki64]{
        \begin{minipage}[b]{0.23\textwidth}
        \includegraphics[width=1\textwidth]{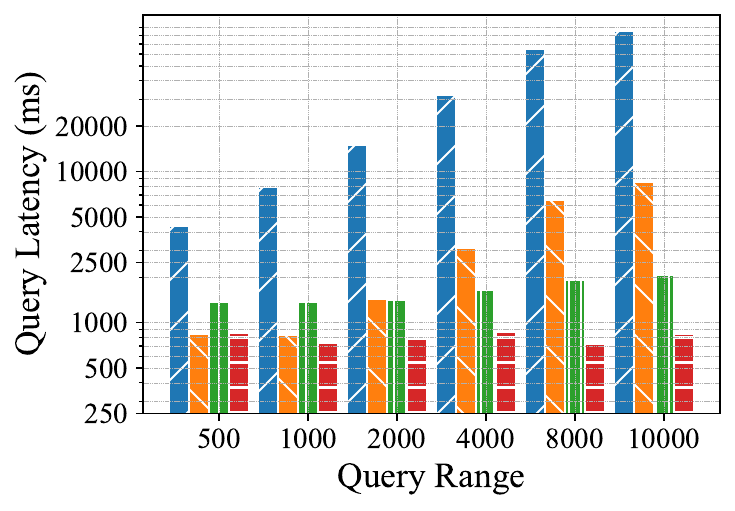}
        \end{minipage}
    \label{sim_r4}
    } \vspace{-1ex}
    \caption{Latency of range queries on various datasets from large-scale simulations}
     \vspace{-3ex}
    \label{sim_r}
\end{figure*}

\begin{figure*}[ht]
    \centering
    \setlength{\subfigcapskip}{-5pt}
    \begin{minipage}[b]{0.99\textwidth}
    \includegraphics[width=1\textwidth]{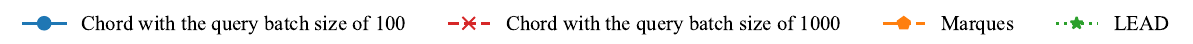} 
    \end{minipage}
    \subfigure[osmc64]{
        \begin{minipage}[b]{0.23\textwidth}
            \includegraphics[width=1\textwidth]{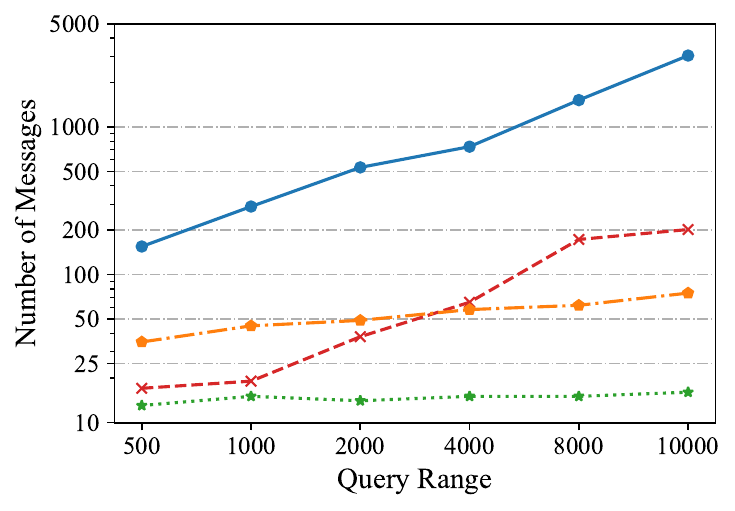} 
        \end{minipage}
        \label{sim_s1}
    }
    \subfigure[face64]{
        \begin{minipage}[b]{0.23\textwidth}
        \includegraphics[width=1\textwidth]{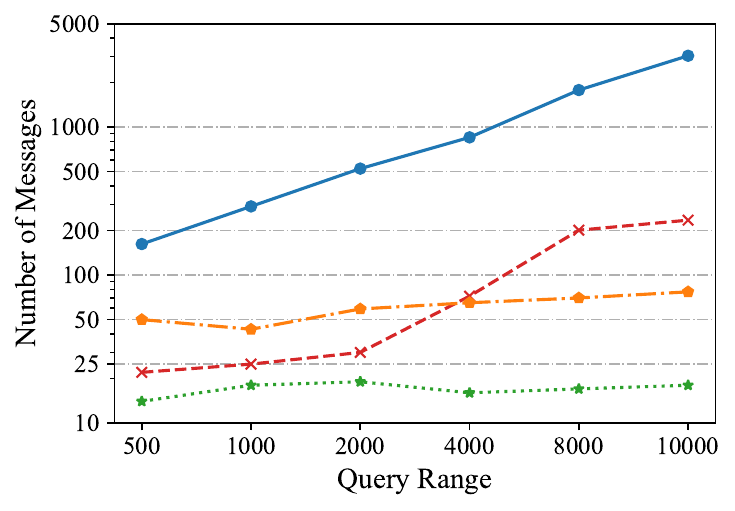}
        \end{minipage}
    \label{sim_s2}
    }
    \subfigure[amzn64]{
        \begin{minipage}[b]{0.23\textwidth}
            \includegraphics[width=1\textwidth]{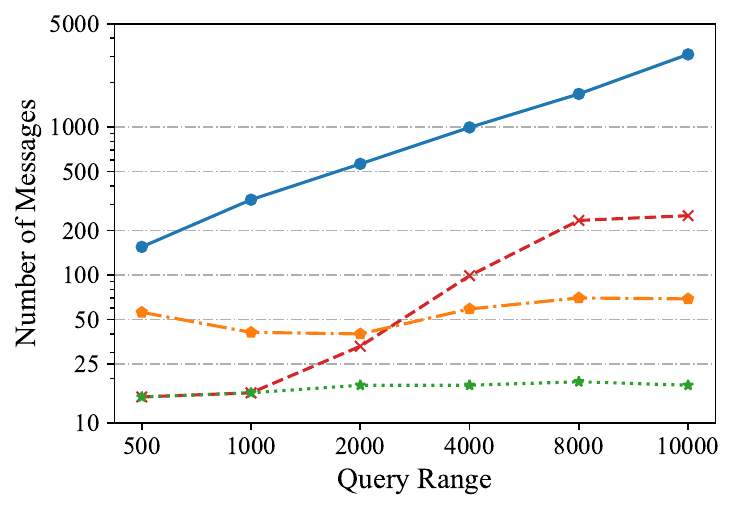} 
        \end{minipage}
        \label{sim_s3}
    }
    \subfigure[wiki64]{
        \begin{minipage}[b]{0.23\textwidth}
        \includegraphics[width=1\textwidth]{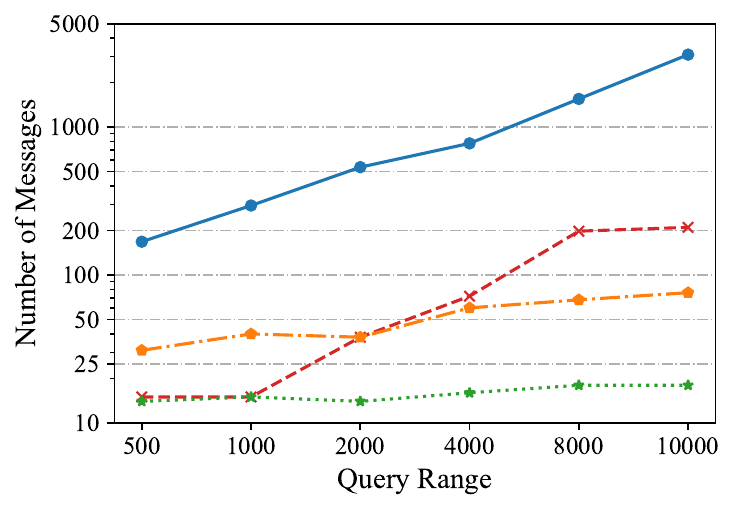}
        \end{minipage}
    \label{sim_s4}
    } \vspace{-1ex}
    \caption{Number of messages of each range query on various datasets.}
     \vspace{-2ex}
    \label{sim_s}
\end{figure*}

\begin{figure*}[htbp]
    \centering
    \setlength{\subfigcapskip}{-3pt}
    \begin{minipage}[b]{0.99\textwidth}
    \includegraphics[width=1\textwidth]{Figures/legend1.pdf} 
    \end{minipage}
    \subfigure[Single Key Lookup]{
        \begin{minipage}[b]{0.23\textwidth}
            \includegraphics[width=1\textwidth]{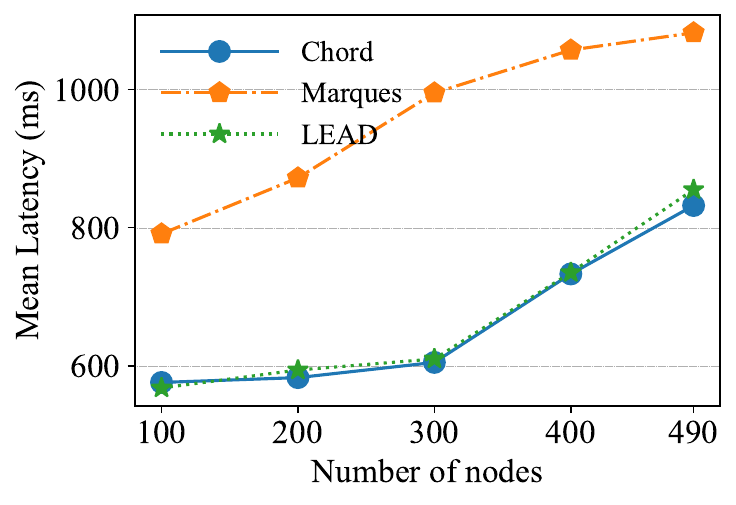}
        \end{minipage}
        \label{singlekey}
    }
    \subfigure[Network Churns]{
        \begin{minipage}[b]{0.23\textwidth}
            \includegraphics[width=1\textwidth]{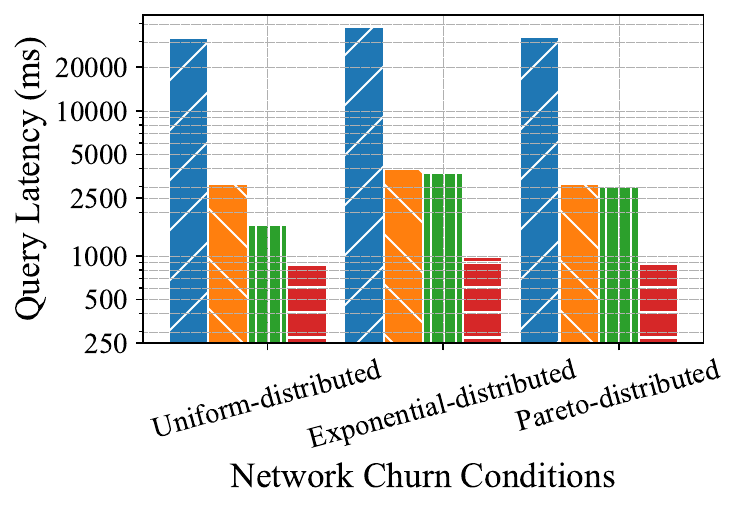}
        \end{minipage}
        \label{churn}
    }
        \subfigure[Network Scales]{
        \begin{minipage}[b]{0.23\textwidth}
        \includegraphics[width=1\textwidth]{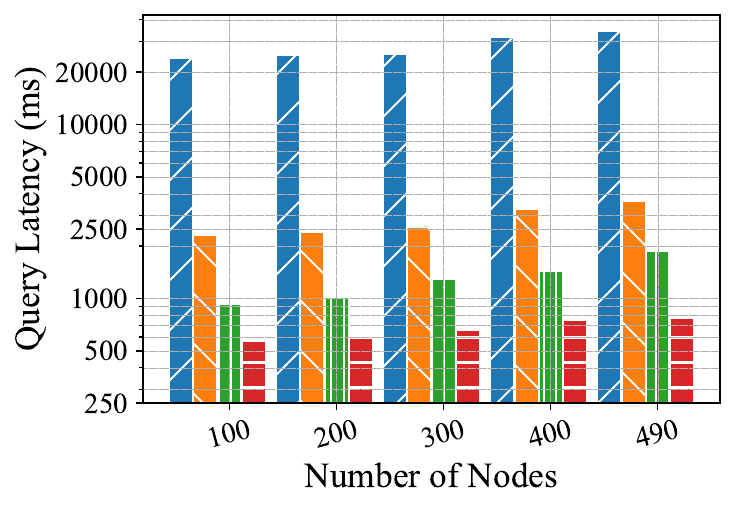}
        \end{minipage}
    \label{nscale}
    }
        \subfigure[Network Topology]{
        \begin{minipage}[b]{0.23\textwidth}
            \includegraphics[width=1\textwidth]{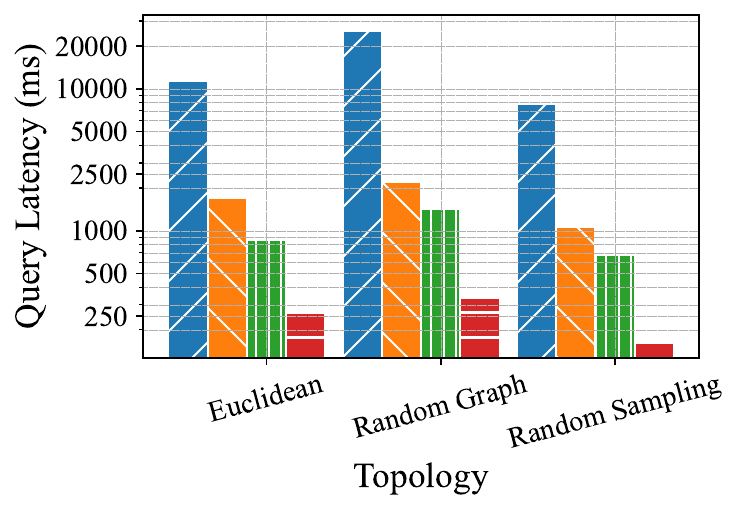} 
        \end{minipage}
        \label{topology}
    }

     \vspace{-1ex}
    \caption{Latency of range queries under various conditions}
     \vspace{-4ex}
    \label{s10}
\end{figure*}
Fig.~\ref{real} presents the latency benchmark results obtained from the real-machine testbed implementation. 
For LEAD, a pre-trained two-layer model incorporating both linear and cubic layers is employed. For each experimental run, we inserted 200 million 64-bit unsigned integers from each of the four datasets. 
Then, we conducted range queries for ranges with varying numbers of keys, from 500 to 10,000, subsequent to a specified key. To ensure the reliability of the results, each query was repeated ten times, and we calculated the average latency for each data point. As demonstrated in Fig.~\ref{real}, as the query range expands, LEAD maintains near-constant latency for range queries. In contrast, both the Batch Query method and Marques exhibit rapidly increasing latencies. For instance, in the experiment using the 'osmc64' dataset, a range query for 500 keys resulted in latencies of 259 ms for Batch Query with a batch size of 1000 and 557 ms for Marques, while LEAD efficiently resolved the query in just 145 ms. As the query range extended to 4,000 keys, the latency for Batch Query escalated to over 1,300 ms and for Marques to over 750 ms. Such latencies become prohibitive for most high-throughput applications. LEAD continued to deliver results in less than 150 ms, showcasing its superior performance and scalability.

\subsection{Simulation Results}\label{Simulation}

The simulated system consists of 490 nodes configured according to the PlanetLab topology. Each node operates 10 virtual nodes. By default, we employed the pretrained two-layer models for LEAD, which incorporates both linear and cubic layers. Each simulation spanned a logical duration of 120 minutes. To emulate the dynamic nature of real-world distributed systems, node lifetimes were modeled with a uniform distribution, averaging 80 logical minutes. The network dynamics were initiated by exiting nodes from the network after their lifespan concluded and rejoining them following a uniformly distributed interval, averaging 10 logical minutes. Each time a node exited and rejoined, its routing state was reset to preserve network integrity. Furthermore, to adapt to network changes effectively, the stabilization timer for each peer was set to 1000 logical ms, enabling regular updates to their finger tables and stabilization of their successor states. Range queries were conducted at regular intervals of five logical minutes throughout the simulation. Each query aimed to retrieve a sequence of $N$ keys subsequent to a specified key $M$. For each query, we documented both the latency and the number of routing steps incurred. Following the completion of each test run, we calculated the average values for these metrics. 

\subsubsection{Range query performance}
Fig. \ref{sim_r} illustrates the range query latency obtained from the simulation. Each experimental cycle involved inserting 200 million 64-bit unsigned integers from one of four distinct datasets: 'osmc64', 'face64', 'amzn64', and 'wiki64'. As the query range extends, the near-constant latency exhibited by LEAD underscores its substantial superiority in query latency compared to other baseline methods across all datasets tested. Again, the results show that LEAD significantly reduces the range query latency.

\subsubsection{Query messages}
To complement our latency analysis, we quantified the number of messages for each range query executed. 
Fig.~\ref{sim_s} depicts the number of messages for range queries required across various test configurations, elucidating LEAD’s optimized path efficiency for range queries. 
LEAD costs much fewer messages compared to the other baselines. 
For example, when the query range is 5000, LEAD only costs $<15$ messages per query, while Marques needs $>50$ messages and Chord, even with batching, requires $>200$ messages for size 1000 and $>1000$ messages for size 100. \textbf{LEAD reduces the query messages by over 80\%.}
We observe that LEAD typically incurs an amount of messages similar to those of a single-key lookup, which is logarithmic relative to network size. 


\subsubsection{Single-key performance}

Alongside evaluating range query performance, we scrutinized the single-key lookup performance of each baseline method, utilizing the `osmc64' dataset as a representative example. Fig. \ref{singlekey} demonstrates LEAD upholds competitive performance with Chord in single-key query latency. This is attributed to its adherence to the foundational design of Chord. On the other hand, Marques's multi-level overlay structure introduces more than a 50\% increase in latency for single-key queries compared to the original Chord.

\subsubsection{Network churn resistance}
As illustrated in Fig. \ref{churn}, the resilience of LEAD is demonstrated through its ability to maintain continuous service performance under various network churn conditions. The test setup involved populating the system with 200 million key-value pairs from the `osmc64' dataset and executing range queries for 4,000 keys. Then we emulated the network dynamics through exiting nodes from the network after their lifespan concluded and rejoining them in intervals that followed uniform, exponential, or Pareto distributions.

\subsubsection{Network scale}
To assess scalability, we varied the network size from 100 to 490 nodes. We utilized the 'osmc64' dataset, which consists of 200 million key-value pairs, to measure latency by executing range queries for 4,000 keys. Fig. \ref{nscale} illustrates that LEAD consistently outperforms other baseline methods across all network sizes evaluated.

\subsubsection{Network topology}
Fig. \ref{topology} demonstrates that LEAD consistently surpasses other baseline methods in range query latency across all evaluated network topologies. In continuation of our scalability testing, with the network size held constant at 490 nodes, we assessed the performance of LEAD across three synthetic network topologies: 1) Euclidean, in which  the latencies between nodes were modeled by their distances in a two-dimensional Euclidean space; 2) Random graph; and 3) Random sampling, in which the inter-node latencies were randomly assigned within a range.

\subsubsection{Load balancing}
\begin{figure}[t]
    \centering
    \setlength{\subfigcapskip}{-6pt}
    \vspace{-1ex}
    \begin{minipage}[b]{0.35\textwidth}
    \includegraphics[width=1\textwidth]{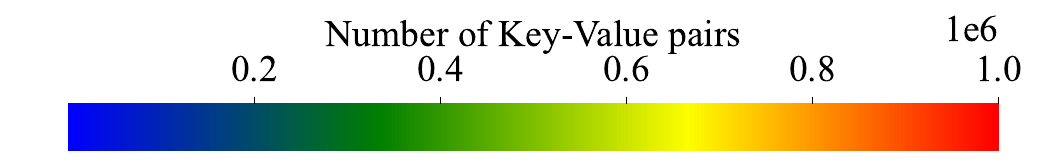} 
    \end{minipage}
    \subfigure[Chord]{
        \begin{minipage}[b]{0.2\textwidth}
            \includegraphics[width=1\textwidth]{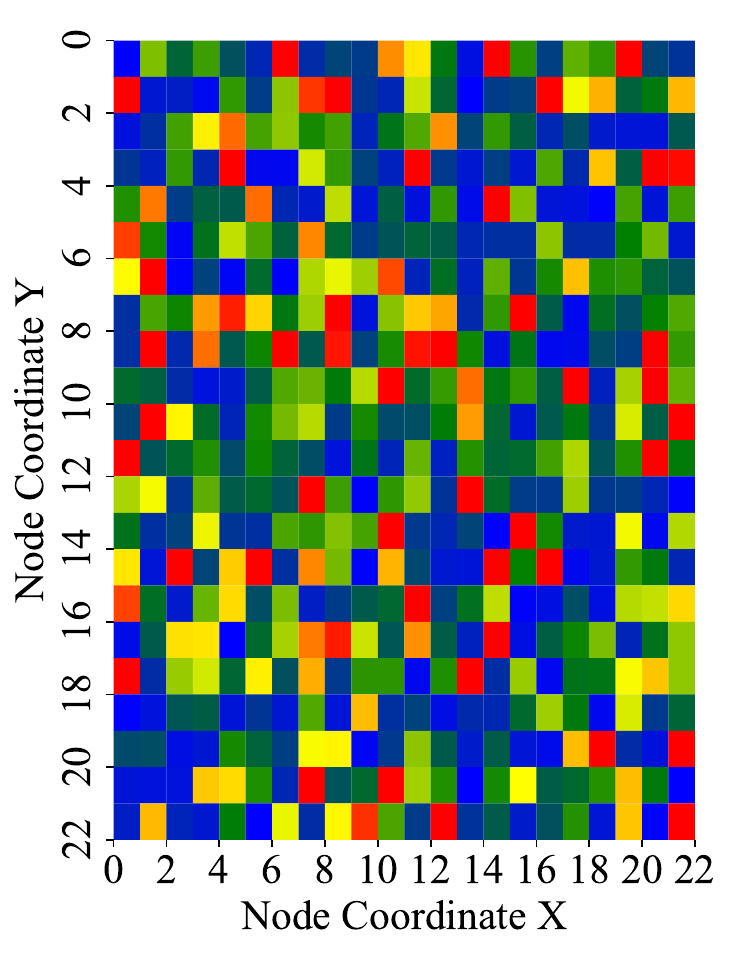} 
        \end{minipage}
        \label{f25}
    }
    \subfigure[LEAD]{
        \begin{minipage}[b]{0.2\textwidth}
        \includegraphics[width=1\textwidth]{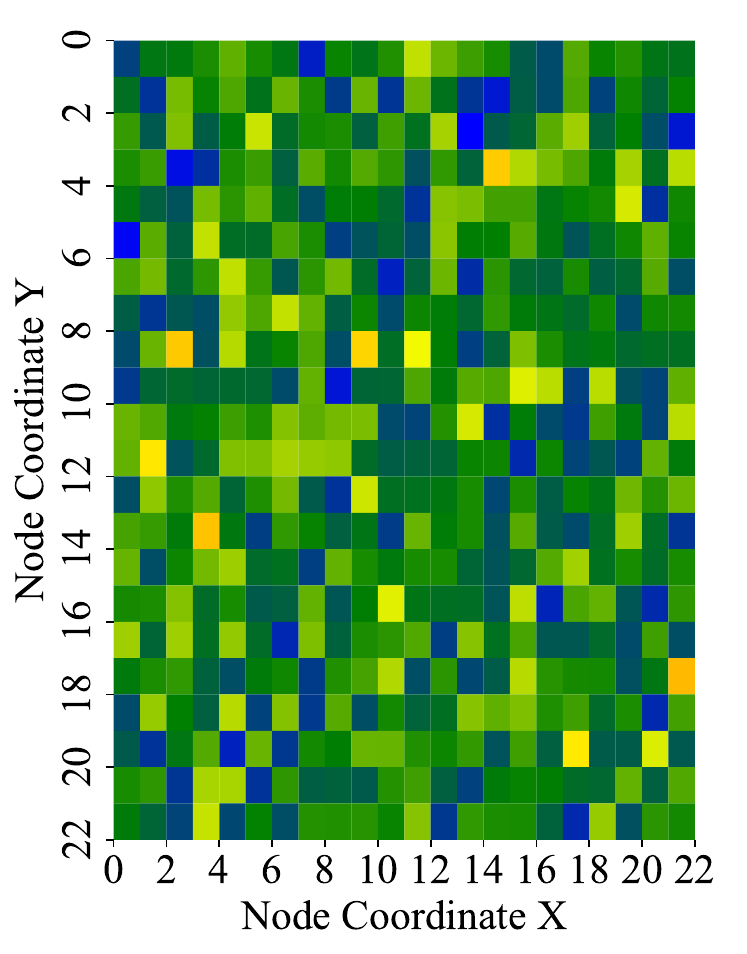}
        \end{minipage}
    \label{f26}
    } \vspace{-2ex}
    \caption{Comparison of Key-Value Pair Distribution}
     \vspace{-4ex}
    \label{load1}
\end{figure}

\begin{figure*}
	\begin{minipage}[t]{0.33\linewidth}
		\centering
		\includegraphics[width=2.25in]{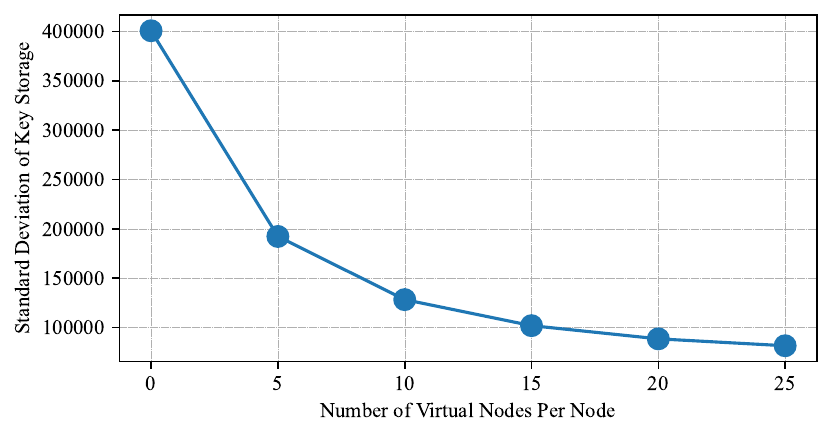}
        \vspace{-3ex}
		\caption{Node Storage Standard Deviation (SD.)}
		\label{f27}
	\end{minipage}
	\begin{minipage}[t]{0.33\linewidth}
		\centering
		\includegraphics[width=2.4in]{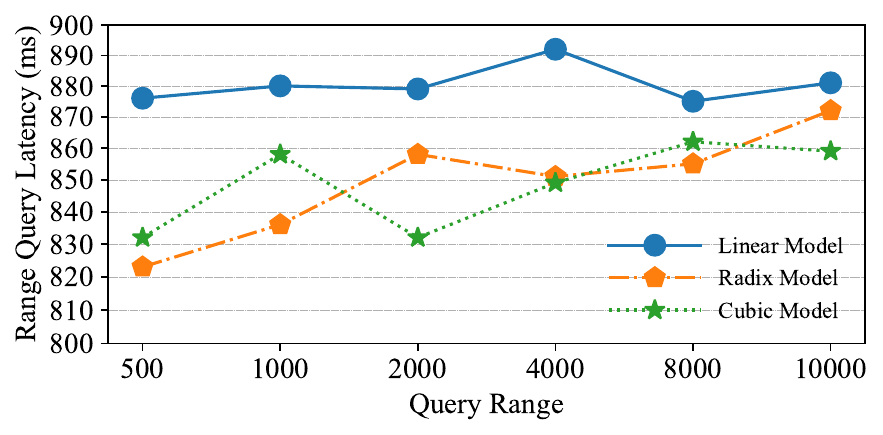}
        \vspace{-6ex}
		\caption{Latency vs. RMI Models}
		\label{model}
	\end{minipage}
    \begin{minipage}[t]{0.33\linewidth}
		\centering
		\includegraphics[width=2.4in]{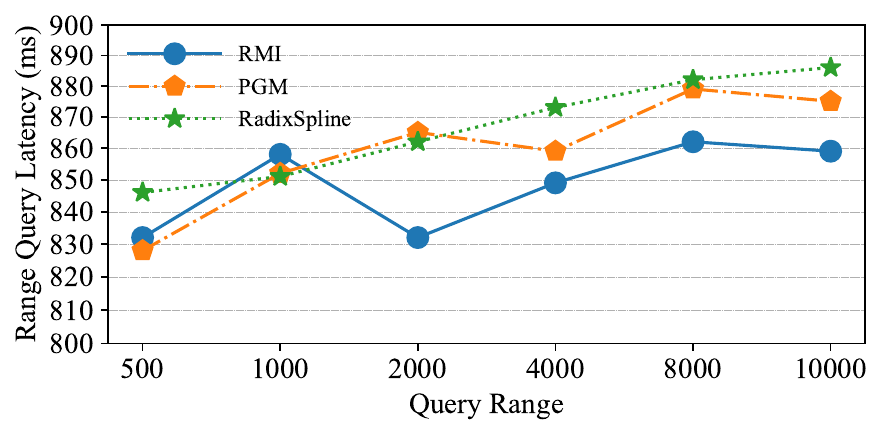}
        \vspace{-6ex}
		\caption{Latency vs Learned Indexes}
		\label{f28}
	\end{minipage}
    \vspace{-5ex}
\end{figure*}

Figure \ref{load1} compares the load distribution of a traditional Chord DHT setup against LEAD integrated with our \textit{Shadow Balancer}, which enables 10 virtual nodes per physical node. The heat map representation shows that LEAD significantly enhances load balancing within the network. 
LEAD with Shadow Balancer exhibits a more uniform green color across the network, indicating a well-balanced load among nodes.
As depicted in Fig.~\ref{f27}, increasing the number of virtual nodes decreases the standard deviation, suggesting better load dispersion. Specifically, the inflection point at 10 virtual nodes per node in the PlanetLab topology indicates an optimized balance. Beyond this point, additional virtual nodes do not significantly improve load balancing, thereby identifying 10 virtual nodes per node as an ideal configuration for the established network.

\subsubsection{Learned models}

\begin{table}[htbp]
\vspace{-5ex}
\caption{Recursive Model Evaluation}
\vspace{-2ex}
\centering
\begin{tabular}{lccc}
\hline
Model  & Maximum Log2 Error & Average Log2 Error & Size (Mb) \\ \hline
Linear & 25.79  & 18.51  & 0.75      \\
Radix  & 21.28  & 12.79  & 1.75      \\
Cubic  & 18.63  & 9.82   & 12.00     \\ \hline
\end{tabular}%
\label{models}
\vspace{-2ex}
\end{table}

LEAD leverages the Recursive Mode structure for fast and accurate order-preserving key mapping with the learned hash function. Selecting the optimal model type during the training phase is crucial to minimize the prediction error, thereby ensuring LEAD achieves optimal load distribution. Fig. \ref{model} illustrates the results of range query latency on the `osmc64' dataset for queries ranging from 500 to 10,000 keys using three typical models as detailed in Table \ref{models}. For the `osmc64' dataset, our evaluations revealed that both the Radix and Cubic models can aptly fit its distribution, showcasing effective performance in managing range queries. In contrast, although the linear model offers benefits in terms of smaller model size, it results in increased error bounds, which can adversely affect the system performance.

\subsubsection{Learned Indexes with LEAD}
\label{sec:models}
In formulating the learned model for LEAD, we assessed various learned index structures, including RMI~\cite{kraska2018case}, Radix Spline Indexes~\cite{10.1145/3401071.3401659}, and Piece-wise Geometric Model Indexes (PGM)~\cite{ferragina2020pgm}. 
Our evaluations, depicted in Figure \ref{f28}, highlight the Recursive Model structure's consistent performance advantage across various query ranges when integrated with LEAD. 
Consequently, the Recursive Model structure is the preferred choice for LEAD, ensuring efficient and accurate range query handling.
\begin{figure*}[htb]
	\begin{minipage}[t]{0.31\linewidth}
		\centering
		\includegraphics[width=\linewidth]{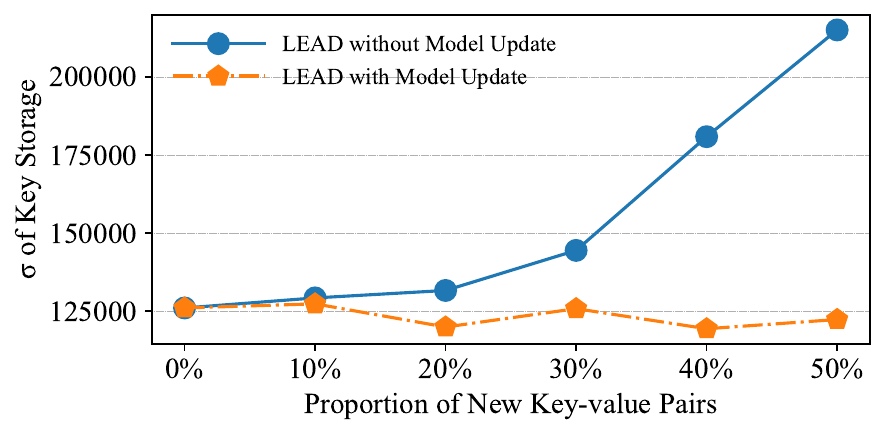}
        \vspace{-5.5ex}
		\caption{SD. of Storage with Model Update}
		\label{f29}
	\end{minipage}
	\begin{minipage}[t]{0.37\linewidth}
		\centering
		\includegraphics[width=\linewidth]{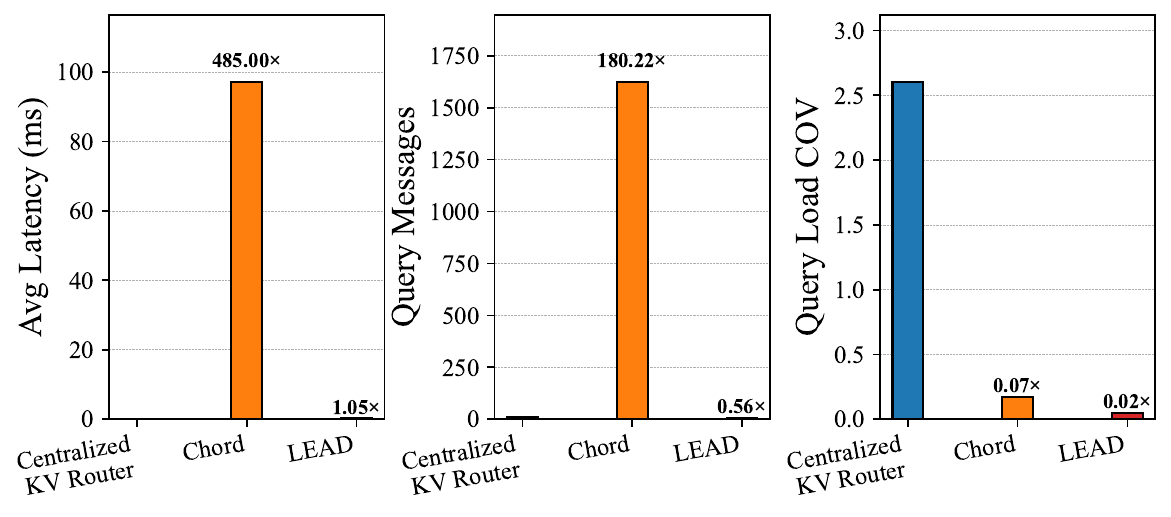}
        \vspace{-5.5ex}
		\caption{KV Cache Management}
		\label{c3}
	\end{minipage}
    \begin{minipage}[t]{0.31\linewidth}
		\centering
		\includegraphics[width=\linewidth]{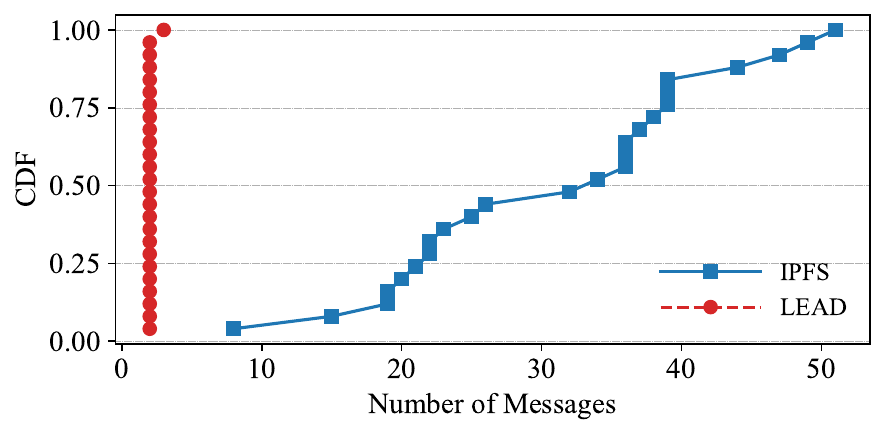}
        \vspace{-5.5ex}
		\caption{CDFs of the message cost in IPFS}
		\label{ipfs}
	\end{minipage}
    \vspace{-5.5ex}
\end{figure*}
\subsubsection{Model Update}
\label{update1}
We randomly selected a portion of the 'osm' dataset as the training set, treating the remainder as unlearned, new key-value pairs. These new pairs were then distributed across the network, alongside the existing data. To quantify the impact of introducing new data on network load balancing, we recorded the standard deviation of key-value pairs stored per node as the proportion of new entries increased. This metric was assessed for both versions of LEAD: one without model updates and one with dynamic model updates enabled. Fig. \ref{f29} illustrates the effects of new key-value pair integration on load balancing across the network. The results indicate that the key mapping with our learned hash function remains stable, with minimum impacts on load balancing, even as new key-value pairs constitute up to 40\% of the network's storage. 
Fig. \ref{f29} also shows that our dynamic model update mechanism sustains optimal load balance across the network, demonstrating LEAD's robust adaptability to data changes within the network.

Our real-world testbed confirms that \textbf{maintaining LEAD’s learned hash function adds only marginal overhead to a vanilla Chord control plane}. At the evaluated scale—one RMI instance per peer collectively managing 200 million 64-bit key–value pairs—the Linear and Radix models occupy under 2 MB of DRAM, while the Cubic model remains below 12 MB. \textbf{Model Update is likewise cheap and invoked only after the system observes a 40 \% drift in new keys}: every key insertion triggers a gradient update that averages 2.1 µs on our testbed—roughly 0.2 \% of end-to-end insertion latency and completely hidden by network delay. The model re-synchronisation exchanging $\leq 60\ \text{KB}$—about one-third of the bytes a  peer already spends during a finger-table refresh cycle. Overall, CPU, memory, and network overheads for the RMI instance stay below 4 \%, 0.1 \%, and 3 \%, respectively.

\subsection{Case study \uppercase\expandafter{\romannumeral1} : KV Cache Management for LLM Severing.}
\label{case3}
In LLM serving, key–value (KV) caches retain the attention keys and values of earlier tokens for reuse rather than recomputation. Hence how to share and re-use existing KV caches is a crucial problem~\cite{vllm,kwon2023efficient,sglang24,srivatsa2024preble,kdn}.
We consider a distributed LLM inference system where multiple nodes (GPU workers) collaboratively serve incoming queries. Each node caches KV blocks~\cite{kwon2023efficient} from sequences it has processed, and nodes cooperate to serve future queries that may need those cached blocks, similar to a CDN. 
The goal is to leverage the existing KV cache instead of recomputing from scratch, by retrieving cached KV blocks from the network. 
KV blocks belonging to the same shared prefix can be stored on the same or nearby nodes. Since an LLM needs the KV for all prior tokens in the sequence, LEAD can fetch a whole span of positions in one efficient range query. That query is routed only to the node(s) responsible for that contiguous key range, obviating any need for system-wide broadcasts or gathers. We experiment a distributed inference system consisting of eight worker nodes—each equipped with a single NVIDIA A100 80 GB GPU running the Llama-3 8B model. The KV cache is managed using the PagedAttention scheme with a fixed 16-token block size~\cite{kwon2023efficient}. Every token is issued a composite key.
Our evaluation workload is Long-DocQA~\cite{li2023loogle}, which comprises 776 lengthy documents paired with 6,400 questions. LooGLE constructs each prompt by prefixing a full document to its associated question; after Zipf-0.6 sampling, prompts average 10,985 tokens~\cite{srivatsa2024preble, fang2025gentorrent}. We simulate a production environment where cold KV blocks can be offloaded to host RAM, resulting in over 500 million KV blocks under management. In the absence of local shared-prefix caching, each inference request must retrieve on average more than 700,000 blocks from the network. We compare LEAD against two baselines: (i) Centralized KV Router: maintains a global index of KV locations across all workers. (ii) Chord-based DHT: uses a distributed hash table overlay for KV storage. Assuming a 10 GbE TCP/IP data-center network with a 100 µs average round-trip latency, we measure each system’s average block-lookup latency, total message count, and cache-hit rate. As Fig. \ref{c3} illustrates, LEAD only add a marginal latency increase to the ideal centralized case, while delivering the resilience and scalability of a fully decentralized design—dramatically outperforming a traditional DHT. Notably, the centralized router suffers from severe query load imbalance, as evidenced by a high query CoV—the coefficient of variation in the number of messages handled per node during range queries. All three systems achieve comparable cache hit rates under identical query correctness guarantees. These results highlight LEAD’s foundation for enabling workload-aware, multi-tiered caching across heterogeneous GPU clusters, supporting dynamic inference pipelines and fault-tolerant, network-aware KV-cache management in scalable LLM serving infrastructures.

\subsection{Case study \uppercase\expandafter{\romannumeral2} : InterPlanetary File System (IPFS).}
\label{case2}
InterPlanetary File System (IPFS)~\cite{trautwein2022design} is a distributed content delivery network that stores, retrieves, and locates data based on the Content Identifiers (CIDs) of its actual content rather than its name or location. With millions of daily content retrievals, IPFS supports numerous third-party applications, demonstrating its broad utility and impact. However, the traditional DHTs used in IPFS, i.e., Kademlia~\cite{maymounkov2002kademlia}, face challenges in handling range queries, which are essential for efficiently retrieving sequences of data blocks or related files. To gauge LEAD’s benefit in a production‑style CDN, we forked the reference go‑ipfs daemon~\cite{ipfscode} and replaced only its routing module with a LEAD overlay, leaving libp2p, Bitswap, and the block‑store unmodified. We emulated an IPFS network with 100 peers using the PlanetLab topology. We generated 100 million synthetic key-value pairs to represent the logical units structuring the metadata of resources (files). For LEAD, the CIDs were managed by the learned hash function within SHA-1’s hashing space. The key operation tested was a typical user request for a resource. In this scenario, an edge server retrieves all the blocks containing the metadata for the requested resource, which consists from 1,000 to 3,000 blocks in our test case. For the emulated IPFS, the server divided the 1,000 lookup queries into 10 batches and sent them over the network. In contrast, for LEAD, the query was optimized as a supported meta-operation range query. We recorded the number of hops required to resolve the request for both systems. As illustrated in Fig. \ref{ipfs}, the CDFs of the retrieval hops for the emulated IPFS and LEAD indicate a significant reduction in the number of messages required to complete data block sequences retrieval when using LEAD.

\section{Related Work}
\label{sec:related}
\textbf{Range query in DHTs.}
Current DHT systems have significant limitations in handling range queries. These systems are inherently designed for exact key-based queries, and therefore, their hashing mechanisms lose the semantic relationship between keys—necessary for range queries. 
Significant efforts to facilitate efficient range queries in distributed networked systems have introduced innovative concepts while also revealing inherent limitations. Early attempts to reconcile hash‑based load‑balancing with ordered access bolted auxiliary data structures onto a vanilla DHT: Prefix‑Hash‑Trees (PHT)~\cite{ramabhadran2004prefix} and Range Search Trees~\cite{gao2004adaptive} layer Chord‐style fingers with a try that must be eagerly split and merged on every insert, leading to high control traffic per update and poor churn tolerance. Armada~\cite{4527242} utilizes a partition tree model and a tailored algorithm within the FissionE~\cite{1498449} topology to enhance range query efficiency. Nevertheless, its reliance on a customized DHT scheme restricts its broader applicability. Similarly, DBST system~\cite{9642540} integrates binary search tree structures to provide efficient range queries for ordered data. 
These tree constructions are assumed to be centralized and are not applicable to large-scale distributed systems. 
MARQUES~\cite{7043516} employs space-filling curves within a multi-level overlay structure derived from Chord~\cite{stoica2001chord}, targeting enhanced performance for range queries. Nonetheless, the complexity involved in managing this structured network overlay can substantially introduce overheads and pose scalability challenges. The latest work, RQIOT~\cite{djellabi2020effective}, tried to employ order-preserving hashing to handle range queries. However, how to implement such a hash method, especially in a dynamic distributed system, is unclear.

\textbf{Learned Index Structures and Hash Functions.}
Recent research has reimagined traditional indexing by conceptualizing indexes as predictive models that estimate the position of a key within a dataset~\cite{kraska2018case,ding2020alex,lu2021apex,wu2021updatable,ferragina2020pgm,tang2020xindex,li2021finedex}. These learned index structures combine machine learning techniques with classical data structures to accelerate key lookups. Kraska et al.~\cite{kraska2018case} proposed the Recursive Model Index (RMI) to address the inaccuracy of using a single model to approximate the dataset’s CDF. There has been growing interest in learned hash functions, where models are trained to map keys to hash buckets in a data-aware manner. Prior works on locality-sensitive hashing (LSH)\cite{wang2017survey,7747793,wang2013order} have explored model-driven hash functions for approximate nearest neighbor search. More recently, Sabek et al.\cite{10.14778/3570690.3570702} demonstrated that learned models can achieve comparable or even fewer hash collisions than traditional hash functions. However, while these approaches show promise, integrating learned index structures or hash functions into decentralized systems such as Distributed Hash Tables remains unexplored—an opportunity that LEAD seeks to address.

\section{Conclusion}
\label{sec:conclude}
This paper introduces LEAD, a novel distributed key-value storage and lookup system designed to enhance the efficiency of range queries by incorporating learned models with DHTs.
LEAD includes the detailed design of training and updating learned models, implementing single-key and range queries, achieving load balancing,  and dealing with system churns. 
Extensive evaluations on both testbed implementation and simulations demonstrate that LEAD significantly reduces the latency and message cost of performing range queries by by $80\%$ to $90\%+$, compared to existing DHT-based solutions.  LEAD can maintain system consistency under dynamic changes and various system conditions. 

We believe LEAD opens a completely new field for further research on integrating learned models with distributed systems. The implementation details are publicly available at \url{https://github.com/ShengzeWang/LEAD}. 
\section*{Acknowledgment}
The authors were partially supported by NSF Grants 2322919, 2420632, 2426031, and 2426940. We thank the anonymous shepherd and reviewers for their valuable comments. 

\bibliographystyle{IEEEtran}
\bibliography{ref}
\clearpage
\appendices

\section*{Appendix}

\subsection{Benchmark Details}
\label{sec:benchmark}
 In our microbenchmark, the Centralized Table (C-Table) baseline represents an approach where a single node maintains a global index mapping key ranges to node locations. This reduces lookup hops but at the cost of storing a huge mapping table on one node – leading to the high memory usage seen in Fig.1(a) – and introducing a single point of failure. The Range-Partition BST (RP-BST) baseline refers to a distributed overlay where nodes are organized in a binary search tree by key ranges. Each node must maintain pointers to tree neighbors (left/right child pointers) and possibly additional routing state, increasing memory overhead and control complexity. IRP-BST reduced messages compared to a plain DHT, but required significantly more memory per node (storing large tree routing tables or interval state) and struggled under churn (due to costly rebalancing of the tree). Fig.1 shows that while these approaches can improve query hops, they exact a high memory/control cost or require central coordination, which motivates the need for a more efficient and fully decentralized solution like LEAD. 
 
A naive strategy to enhance range queries is parallelizing lookups on traditional DHTs by simultaneously issuing multiple key queries. The "Chord with batch size 100/1000" baselines represent scenarios where 100 or 1000 parallel lookups are issued concurrently. However, this parallelization introduces substantial practical overhead: (i) The initiating node must manage numerous simultaneous responses, potentially saturating its network interface or CPU. (ii) The overall query latency is still constrained by the slowest individual lookup, limiting scalability and efficiency in realistic conditions.
 
\subsection{Shadow Balancer Analysis}
\label{sec:balancer_analysis}
Consider a LEAD deployment with $n$ physical nodes, each hosting $k$ virtual nodes. The Shadow Balancer introduces additional message complexity proportional only to k, resulting in an overall complexity still logarithmic with respect to n. Experimentally, we found that using 10 virtual nodes per physical node incurred minimal overhead—less than 4\% CPU usage, under 0.1\% additional memory per node for routing tables, and negligible network impact—while significantly enhancing load balancing. This approach aligns with established industry practices; for instance, Apache Cassandra recommends a similar virtual-node strategy to evenly distribute load with minimal overhead. 

Consider a LEAD network composed of $n$ nodes, each of which is virtualized into $m$ virtual nodes. The system is designed to accommodate a hashing space capable of handling $h$ hash values. Each virtual node oversees a virtual finger table containing $b = \lfloor \log h \rfloor$ entries. When a peer $P_0$ with a virtual node ID $H_{P_0}$ initiates a range query for $n$ successive key-value pairs starting from key $K$ with a hash value $H_{K}$, the query will reach $S$ within $\log(mn)$ hops with high probability. With $V$ key-value pairs managed in the network, each peer manages an average of $\frac{V}{mn}$ key-value pairs. Upon reaching $S$, the query retrieves an average of $\frac{V}{2mn}$ keys. If the query is not completed, it continues on $S$'s successors, each of which retrieves $\frac{V}{mn}$ keys. Consequently, after reaching $S$, the query needs an additional $\lceil \frac{mn^2}{V} - \frac{1}{2} \rceil$ hops with high probability. Given that $n^2 \ll V$ and $m \ll V$, it follows that $\frac{mn^2}{V} < 1$ with high probability. Therefore, the range query requires less than one additional hop after reaching $S$, and it can be resolved within $\log(mn)$ hops with high probability.

Furthermore, our design naturally accommodates heterogeneous environments by assigning more virtual nodes (thus greater hash space) to more capable nodes, while assigning fewer to resource-constrained ones. We acknowledge that “hot keys,” receiving disproportionately high request volumes, constitute a distinct load-balancing concern at the request level. Currently, LEAD does not explicitly handle hot-key replication or caching, but such strategies can be integrated orthogonally, for example, via application-level replication. We plan to extend LEAD with standard replication strategies for hot keys in future releases.

\subsection{Handling New Keys and Failure Recovery}
\label{sec:recovery}
LEAD is designed so that correctness is never compromised by model error. Even if new keys arrive that the model hasn’t seen, those keys are still inserted and found correctly. This is because the system always uses the current learned model output in combination with the DHT’s finger tables to place and locate keys. In the worst case, an outdated or inaccurate model could result in uneven load distribution, but the correctness of lookups remains guaranteed by fundamental DHT routing principles. Our evaluation showed that the learned model remained efficient even as up to 40\% new keys were added; beyond that, we triggered a model update to realign with the new CDF. By following successors until N keys are gathered, the query inherently stops only after covering the continuous range requested. As long as the keys are stored in sorted order across the ring, which our learned hash function aims to ensure, order consistency is maintained. If the key distribution is highly irregular or adversarial, a learned model might not yield much benefit. In such a scenario, the performance of LEAD would gradually revert toward the baseline DHT: even if the model’s predictions are poor, the system will still find keys correctly using DHT routing. Our evaluations included diverse real-world datasets — uniform-like distributions, highly skewed popularity data, temporal data, etc. — and LEAD handled all with strong performance gains.

LEAD is designed to adapt rapidly to system churn through robust stabilization and failure recovery mechanisms. The correctness of LEAD is dependent on the current knowledge of its successors and predecessors within the network. Additionally, the efficiency of query handling is contingent upon the timeliness and accuracy of the virtual finger tables. To maintain this information, each peer periodically stabilizes itself in the network through the following mechanisms:

\begin{itemize}
\item[(1)] \textbf{Successor and Predecessor Verification:} Peers regularly initiate verification requests to update their immediate successors and predecessors.
\item[(2)] \textbf{Heartbeat Communications:} Regular heartbeat messages are exchanged between a peer and its network neighbors on their successor and predecessor lists to affirm their presence and operational status.
\item[(3)] \textbf{Virtual Finger Table Maintenance:} Each peer undertakes systematic verification and updates of its virtual finger table to align with the current network state.
\end{itemize}

Each peer continuously monitors the operational status of its immediate neighbors through regular heartbeat messages. A missed series of heartbeats triggers an immediate suspicion of peer failure, prompting further verification actions. After a given timeout, the peer flags it as a failure and initiates the recovery process. In addition to maintaining a primary successor and predecessor, LEAD implements a successor list and a predecessor list containing several backup peers. Upon detection of a failure in the immediate successor, the next available peer from this list is promoted to take over as the immediate successor or predecessor. Concurrently, the peer waits for the next stabilization phase to verify the updated correctness of the immediate successor and predecessor. At the same time, its virtual finger table is updated seamlessly with the successor recovery. Meanwhile, the node managing the virtual peers regularly checks their liveliness as a higher-level observer. Should a virtual peer fail to reestablish its status within the network following specified timeouts, the node initiates a controlled removal of the peer from the network and facilitates the rejoining of the peer via its other operational virtual peers. Furthermore, while the core focus of this paper is to delineate the fundamental operations of LEAD, the implementation of data redundancy and replication can be enabled from a higher level, providing an additional layer of data protection.

\textbf{Correctness guarantee under asynchronous updates.} LEAD preserves query‑correctness despite (i) concurrent model updates, (ii) finger‑table maintenance, and (iii) churn, because these three mechanisms are decoupled and monotone:
\begin{itemize}
\item[(1)] \textbf{Decoupled address spaces:} Ownership is defined solely by the $PeerHASH$, never by the evolving learned model. Even if two peers temporarily disagree on the newest FRM version, they still agree on who ultimately owns any hash identifier.
\item[(2)] \textbf{Version‑monotone models:} Each peer tags its leaf‑model blob with a monotonically increasing version vector; routing messages carry the sender’s current version. A peer never rolls back to an older version, so once all predecessors of an identifier $h$ adopt version $v$, no later update can map $K$ outside that predecessor interval.
\item[(3)] \textbf{Safe forwarding rule:} When a query for $K$ reaches a peer $P$ whose local RMI predicts a different successor than its finger table, P forwards the query along the finger‑table edge—which, by Chord’s proven invariants, always advances the query at least halfway to the true owner. Thus every hop strictly decreases the identifier distance in the canonical ring metric, guaranteeing convergence in $\leq \lceil \log_2 N \rceil$ hops even while RMI versions are in flux.
\item[(4)] \textbf{Eventual convergence:} The heartbeat messages ensure that any live peer receives strictly newer RMI versions from at least one neighbor within $\Delta \leq O(\log N)$ message delays, hence the overlay becomes lookup‑correct again after $O(\log N)$ rounds following the last FRM update.

\end{itemize}

From the perspective of an active peer, there may be occasions where it attempts to route a query through a peer that has failed but is still considered alive prior to the completion of the stabilization process. In such cases, the query will either proceed after a timeout -- prompting the peer to retry the query -- or it may continue through a suboptimal path during this particular hop. Despite these interruptions, the system is designed to stabilize rapidly. This ensures that even a non-optimal routing hop does not compromise the logarithmic routing step guarantee for data retrieval, provided that the overall integrity and correctness of the system are maintained.

\subsection{Fully Decentralized Model Update}
LEAD’s federated update mechanism is designed to be lightweight and distributed. The term “coordinator” used in the paper does not imply a persistent leader or central node; rather, rather, it describes a transient role adopted temporarily by a peer to merge local model updates within its immediate neighborhood. This design ensures that coordinators communicate only with a small subset of nodes, thereby limiting communication overhead and preventing any single node from becoming a global bottleneck. Additionally, as detailed in Section III.E, multiple coordinators can concurrently aggregate updates in different network regions. This parallel and localized process ensures that the model update mechanism scales gracefully. The FRM within the LEAD protocol provides a robust mechanism for decentralized and cooperative model updates among peers, aimed at optimizing load balancing when a significant influx of new key-value pairs is introduced to the network.

\subsection{Security Considerations}
While a full adversarial analysis was outside our original scope, we acknowledge that the learned model could be a potential attack surface if an adversary controls some peers. Importantly, since peer addressing and routing rely on cryptographic hashing independent of the learned model, attackers cannot misroute queries, although they could attempt to degrade load balancing. To mitigate risks, we already utilize gradient aggregation during model updates to prevent any single peer from dominating the training phase and manipulating data availability. In future iterations, we plan to collaborate with the broader community to further harden LEAD by incorporating established security mechanisms such as Sybil-resistant identities and robust aggregation techniques to counteract malicious nodes.

\subsection{Additional Evaluation Results}


\subsubsection{Case study \uppercase\expandafter{\romannumeral3}: Blockchain application}
\label{case3}
\begin{figure}[htb]
	
		\centering
		\includegraphics[width=\linewidth]{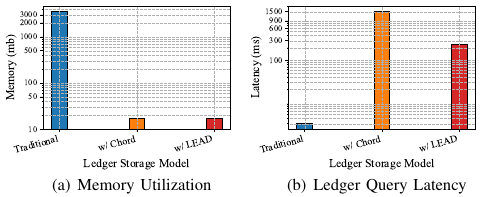}
        \vspace{-5ex}
		\caption{Range Query Benchmark in Blockchain}
		\label{block}
\end{figure}
We perform a case study to show how LEAD can improve data retrieval performance in practice for blockchain applications. Specifically, we apply LEAD on a distributed collaborative ledger system~\cite{nie2024collaborative}. Blockchain applications, particularly those involving collaborative ledgers, require efficient and reliable data access to ensure consistency and speed in transaction processing. For example, verification tasks~\cite{MARIJAN2022100492} often require the retrieval of specific ranges of blocks or transactions to confirm their validity, ensure they adhere to the chain rules, or perform security checks. Traditional DHT implementations provide a robust structure for single key queries but often struggle with latency issues as the query range scales up due to their reliance on naive mechanisms that lack support for efficient range queries. LEAD can be seamlessly integrated into blockchain systems to alleviate bottlenecks associated with range query latency. 
In this case study, we demonstrate this by substituting the traditional DHTs used in the ledger storage system~\cite{nie2024collaborative} with LEAD. We loaded the system with a blockchain ledger containing 100 million organic timestamps as keys linked to their corresponding transaction values. Subsequently, we conducted range queries for 4,000 transactions following a specific timestamp. We compared LEAD with two other baseline methods: the Chord collaborative ledger storage model as proposed by \cite{nie2024collaborative} and a naive full ledger storage model that maintains a complete copy of the ledger on each peer, as depicted in Fig. \ref{block}. The naive full ledger storage model, while offering minimal in-memory query latency due to its localized data access, incurs substantial memory overhead by storing the entire ledger. This approach, although straightforward, becomes impractical as the ledger size grows, leading to scalability issues. In contrast, LEAD demonstrates a significant improvement in reducing ledger query latency while concurrently optimizing memory usage when compared to the traditional DHT storage model.

\subsection{Additional Related Work}
\textbf{Learning Hash Functions.} 
To date, the concept of the "Learned Hash Function" has not been formally established in the literature. Nonetheless, there have been attempts to develop data-dependent hash functions that learn from the characteristics of the data itself. For instance, research such as~\cite{wang2017survey, 7747793, wang2013order} has explored learning locality-sensitive hash (LSH) functions to construct Approximate Nearest Neighborhood (ANN) indexes. These studies are primarily focused on clustering similar items into buckets to facilitate nearest neighborhood queries, yet their methodologies are not adaptable to a broader range of applications~\cite{kraska2018case, richter2015seven, wang2017survey}. Additionally,  a recent study by Sabek et al.\cite{10.14778/3570690.3570702} discusses the potential to replace traditional hash functions with learned models. It indicates that learned models produce a similar or reduced number of hash collisions compared to traditional hash functions. There has been no exploration into integrating learned index structures within distributed networked systems yet.

\textbf{KV Cache Management for LLM Severing.} 
Inference with large language models over extended contexts incurs substantial memory and latency overhead because self-attention scales quadratically with sequence length. To mitigate this, key–value (KV) caching retains the attention keys and values of earlier tokens for reuse rather than recomputation, trading additional memory for reduced compute—an especially worthwhile compromise when long shared prefixes would otherwise demand exorbitant recomputation~\cite{shi2024keep, kwon2023efficient, yao2025cacheblend}. However, because the KV cache grows linearly with sequence length, it can quickly exhaust a single node’s GPU memory , and coordinating caches across multiple nodes introduces considerable data-management overhead.
In LLM serving, key–value (KV) caches retain the attention keys and values of earlier tokens for reuse rather than recomputation. Hence how to share and re-use existing KV caches is a crucial problem~\cite{vllm,kwon2023efficient,sglang24,srivatsa2024preble,kdn}. Our case study \uppercase\expandafter{\romannumeral1} is analogous to a content delivery network (CDN), where each node can serve cache lookups for others. We assume a realistic scenario of long-context reuse: queries arrive in some temporal order (here following a Zipf-0.6 popularity distribution in our evaluation), and if the shared prefix was recently used on one node, another query for the same document might land on a different node. 
The goal is to leverage the existing KV cache instead of recomputing from scratch, by retrieving cached KV blocks from the network. LEAD can organize the KV cache across nodes to optimize such reuse. Rather than hashing these KV blocks arbitrarily across the system, LEAD maintains contiguous ranges of keys. 

\subsection{Expanding LEAD into Emerging Domains}
LEAD opens several avenues for further research. The architectural principles behind LEAD—order‑preserving learned hashing, cooperative model updates, and lightweight virtual‑node balancing—extend well beyond classical key‑value stores. We outline several promising, high‑impact directions that we plan to prototype:
\subsubsection{Vector Databases and Retrieval‑Augmented Generation}
Vector search engines such as Milvus~\cite{wang2021milvus}, Pinecone~\cite{Pinecone}, and Meta’s FAISS~\cite{faiss} shard billions of embeddings but rely on static partitioners or HNSW graphs that must be rebuilt as data drift. By hashing product‑quantization or IVF bucket IDs through LEAD, semantically adjacent vectors would land on neighboring peers, enabling locality‑aware ANN probes while retaining the high-efficient routing guarantee. Because the CDF is retrained cooperatively, the overlay can track concept drift without global re‑indexing—an attractive property for Retrieval‑Augmented‑Generation (RAG)~\cite{lewis2020retrieval} stacks that ingest fresh documents continuously. Early work on peer‑to‑peer vector search underscores the demand for such decentralized indexes~\cite{pan2024vector}.
\subsubsection{Large‑Scale LLM Serving and Decentralized Inference}
Looking forward, we see LEAD as a compelling substrate for next‑generation AI serving stacks, especially for sharding the ever‑growing KV caches and model adapters that dominate large‑language‑model (LLM) inference. By hashing tuples ⟨sequence‑ID,layer,token‑pos⟩ through LEAD’s order‑preserving learned function, temporally adjacent tokens land on neighbouring peers, enabling sliding‑window eviction and locality‑aware reuse without a central KV router—an attractive alternative to recent distributed‑attention proposals such as DistAttention~\cite{lin2024infinite}, KV‑Runahead~\cite{cho2024kv}, PRESERVE~\cite{yuzuguler2025preserve}.

\subsubsection{Edge‑AI and IoT Data Lakes}
Massive edge deployments generate time‑series telemetry that must be queried by geographic window or recency~\cite{singh2023edge, shi2020communication}. LEAD can hash composite keys ⟨device‑ID, timestamp⟩ so that adjacent time ranges map to adjacent roadside or gateway peers. The learned CDF absorbs workload skew caused by bursty sensors, while FRM keeps models current without cloud coordination—ideal for bandwidth‑constrained, privacy‑sensitive edge fabric.
\subsubsection{Content‑Delivery Networks and caches}
Modern CDNs already rely on consistent hashing to place objects across Point of Presences (PoPs)~\cite{passarella2012survey}, but these static hashes ignore temporal popularity skew, leading to cache imbalance and cold‑start misses. Integrating LEAD at the cache‑mapping layer would allow the overlay to learn the content popularity CDF and adaptively remap hot prefixes to additional PoPs, reducing origin fetches. This direction aligns with the long‑standing goal of “demand‑responsive” CDNs envisioned by CDN providers~\cite{Akamai}.
\subsubsection{Cross‑chain and indexing}
Inter‑ledger stacks need fast, ordered look‑ups across heterogeneous blockchains~\cite{zheng2018blockchain,monrat2019survey}. A LEAD overlay per chain—bridged via light‑client proofs—would supply uniform range queries over block‑height or asset‑ID without a trusted indexer, complementing recent cross‑chain initiatives.
\subsubsection{Harden LEAD against adversarial behaviour}
A promising complementary effort is to harden LEAD against adversarial behaviour. Because the learned hash derives from live data, attackers might (i) inject skewed keys to distort the CDF, (ii) return bogus model parameters to misroute queries, or (iii) join the overlay with large Sybil swarms. We plan to explore lightweight counter‑measures—replicated model cross‑checks, capped learning rates, signed model digests, and standard Sybil‑throttling rules—that preserve LEAD’s efficiency while adding strong resilience.

Implementation details, as well as simulation code, are publicly available at \url{https://github.com/ShengzeWang/LEAD}. We intend to release LEAD as a robust, accessible library, facilitating continued research within the community and encouraging broader industry adoption.

\end{document}